\documentclass[final,authoryear,3p,times]{elsarticle}

\usepackage{afterpage}
\usepackage{algorithm}
\usepackage{algpseudocode}
\usepackage{amsmath}
\usepackage{amssymb}
\usepackage{amsthm}
\usepackage{bm}
\usepackage{booktabs}
\usepackage{calc}
\usepackage{caption}
\usepackage{subcaption}
\usepackage{color}
\usepackage{colortbl}
\usepackage[dvipsnames]{xcolor}
\usepackage{enumerate}
\usepackage{graphicx}
\usepackage{ifthen}
\usepackage[parfill]{parskip} 
\usepackage{multicol}
\usepackage{siunitx}
\usepackage{ulem}

\usepackage[colorlinks]{hyperref}
\usepackage[nameinlink,capitalise]{cleveref}
\crefname{appendix}{}{}

\usepackage{tikz}
\usetikzlibrary{arrows}
\usetikzlibrary{automata}
\usetikzlibrary{calc}
\usetikzlibrary{chains}
\usetikzlibrary{positioning}
\usetikzlibrary{shapes}

\usepackage{varwidth}

\usepackage{tcolorbox}
\newtcolorbox[auto counter]{problem}[2][]{colframe=blue!30, colback=blue!5, coltitle=black, title=Problem~\thetcbcounter~ #2,#1}

\usepackage{spverbatim} 
\usepackage{listings} 

\edef\svtheparindent{\the\parindent}
\usepackage{parskip}
\parindent=\svtheparindent\relax

\newcommand{\LASSO}{LASSO}
\newcommand{\LARS}{LARS}
\newcommand{\LARSLASSO}{LARS-LASSO}

\definecolor{ForestGreen}{RGB}{34,139,34}
\definecolor{InternationalOrange}{rgb}{1.0, 0.31, 0.0}
\definecolor{WineRed}{RGB}{139,0,0}

\newcommand{\refp}[1]{Problem~\ref{#1}}
\newcommand{\boldface}[1]{\boldsymbol{#1}}  

\newcommand{\bfb}{\boldface{b}}
\newcommand{\bfc}{\boldface{c}}
\newcommand{\bfd}{\boldface{d}}

\newcommand{\bfn}{\boldface{n}}

\newcommand{\bfr}{\boldface{r}}
\newcommand{\bfs}{\boldface{s}}

\newcommand{\bfu}{\boldface{u}}

\newcommand{\bfw}{\boldface{w}}

\newcommand{\bfy}{\boldface{y}}

\newcommand{\bfC}{\boldface{C}}

\newcommand{\bfF}{\boldface{F}}

\newcommand{\bfI}{\boldface{I}}

\newcommand{\bfN}{\boldface{N}}

\newcommand{\bfP}{\boldface{P}}
\newcommand{\bfQ}{\boldface{Q}}
\newcommand{\bfR}{\boldface{R}}

\newcommand{\bfU}{\boldface{U}}

\newcommand{\bfX}{\boldface{X}}

\newcommand{\bfdelta}{\boldsymbol{\delta}}

\newcommand{\bfmu}{\boldsymbol{\mu}}

\newcommand{\Rset}{\mathbb{R}}

\DeclareMathOperator{\diag}{diag}

\DeclareMathOperator{\sign}{sign}

\newcommand{\argmax}{\operatornamewithlimits{arg\ max}}
\newcommand{\argmin}{\operatornamewithlimits{arg\ min}}

\newcommand{\be}{\begin{equation}}
\newcommand{\ee}{\end{equation}}
\newcommand{\bea}{\begin{equation}\begin{aligned}}
\newcommand{\eea}{\end{aligned}\end{equation}}
\newcommand{\beq}{\begin{eqnarray}}
\newcommand{\eeq}{\end{eqnarray}}
\newcommand{\bem}{\begin{multline}}
\newcommand{\eem}{\end{multline}}
\newcommand{\ba}{\begin{align}}
\newcommand{\ea}{\end{align}}
\newcommand{\bcase}{\left\{ \begin{array}{ll}}
\newcommand{\ecase}{\end{array} \right.}
\newcommand{\CHANGE}[1]{\textcolor{black}{#1}}

\begin{document}

\begin{frontmatter}

\title{Non-smooth optimization meets automated material model discovery}

\author[fau,stan]{Moritz Flaschel\corref{cor1}}
\author[stat]{Trevor Hastie}
\author[fau,stan]{Ellen Kuhl}

\cortext[cor1]{Correspondence: moritz.flaschel@fau.de}

\address[fau]{Institute of Applied Mechanics, Egerlandstraße 5, Friedrich-Alexander-Universität Erlangen–Nürnberg, 91058 Erlangen, Germany}
\address[stan]{Department of Mechanical Engineering, Stanford University, 440 Escondido Mall, California 94305, United States.}
\address[stat]{Department of Statistics, Stanford University, Sequoia Hall 390 Jane Stanford Way, California 94305-4065, United States.}

\begin{abstract}
Automated material model discovery has gained significant traction in recent years, as it disrupts the tedious and time-consuming cycle of iteratively calibrating and modifying manually designed models.
Non-smooth $L_1$-norm regularization is the backbone of automated model discovery;
however, the current literature on automated material model discovery offers limited insights into the robust and efficient minimization of non-smooth objective functions.
In this work, we examine the minimization of functions of the form $f(\bfw) + \alpha \|\bfw\|_1$, where $\bfw$ are the material model parameters, \mbox{$f$ is} a metric that quantifies the mismatch between the material model and the observed data, and \mbox{$\alpha \geq 0$} is a regularization parameter that determines the sparsity of the solution.
We investigate both the straightforward case where $f$ is quadratic and the more complex scenario where it is non-quadratic or even non-convex.
\CHANGE{Importantly, in contrast to previous works on automated material model discovery, we do not only focus on methods that solve the sparse regression problem for a given value of the regularization parameter $\alpha$, but propose methods to efficiently compute the entire regularization path, facilitating the selection of a suitable $\alpha$.}
Specifically, we present four algorithms and discuss their roles for automated material model discovery in mechanics:
First, we recapitulate a well-known \textit{coordinate descent} algorithm that solves the minimization problem assuming that \mbox{$f$ is} quadratic for a given value of $\alpha$, also known as the \LASSO{}.
Second, we discuss the algorithm \LARS{}, which automatically determines the critical values of $\alpha$, at which material parameters in $\bfw$ are set to zero.
\CHANGE{Third, we propose to use the \textit{proximal gradient method} ISTA for automated material model discovery if \mbox{$f$ is} not quadratic, and fourth, we suggest a novel pathwise extension of ISTA for computing the regularization path.}
\CHANGE{Many of these algorithms have not yet been applied to problems in computational material mechanics.}
We demonstrate the applicability of all algorithms for the automated discovery of incompressible hyperelastic material models from uniaxial tension and simple shear data.
\end{abstract}


\end{frontmatter}

\section{Introduction}
\label{sec:Introduction}

Traditional material modeling, that is, the manual design of a material model and the calibration of its material parameters, is known to be prone to modeling errors, for example, due to incorrect modeling assumptions or an inappropriate choice of functions describing material behavior.
Consequently, current research focuses on machine learning material models \citep{ghaboussi_knowledgebased_1991,sussman_model_2009,vlassis_geometric_2020,masi_thermodynamics-based_2021,linka_constitutive_2021,bonatti_one_2021,klein_polyconvex_2022,asad_mechanics-informed_2022,fuhg_local_2022,tac_data-driven_2022,thakolkaran_nn-euclid_2022,kalina_automated_2022,rosenkranz_comparative_2023,benady_unsupervised_2024,flaschel_convex_2025,bleyer_learning_2025}, bypassing the formulation of material models in the classical sense \citep{kirchdoerfer_data-driven_2016,ibanez_data-driven_2017}, or automatically discovering material models as interpretable symbolic expressions \citep{schoenauer_evolutionary_1996,ratle_grammar-guided_2001,versino_data_2017,flaschel_unsupervised_2021,bomarito_development_2021,park_multiscale_2021,wang_inference_2021,wang_establish_2022,abdusalamov_automatic_2023,linka_new_2023,meyer_thermodynamically_2023,fuhg_extreme_2024,hou_automated_2024,bahmani_physicsconstrained_2024,kissas_language_2024,thakolkaran_can_2025,abdolazizi_constitutive_2025}, see \cite{fuhg_review_2024} for a comprehensive review.

While each data-driven material modeling method has its merits and is well suited to specific use cases, automatically discovering closed-form mathematical expressions for the material model offers several advantages.
First, material models encoded in concise formulas are memory efficient because they compress all information about the material's behavior into short mathematical expressions with only a few parameters.
Storing a handful of parameters requires less memory than storing the weights of a neural network or a database of stress-strain pairs, as needed for model-free approaches \citep{kirchdoerfer_data-driven_2016,ibanez_data-driven_2017}.
In addition, concise material models are typically more efficient to evaluate compared to other machine learning approaches in which information has to pass, for example, through several layers of a neural network.
This means that concise models naturally lend themselves to finite element simulations \citep{peirlinck_democratizing_2024}.
Finally, automated material model discovery facilitates the physical interpretation of the discovered material behavior and simplifies the communication of the discovered models to other researchers.
With automated material model discovery, we can identify the most suitable functions to describe the material behavior, determine the number of internal variables needed to capture its path-dependence, and automatically classify the material into an appropriate category, such as elasticity, viscoelasticity, or plasticity \citep{flaschel_automated_2023}.

One of the most popular approaches to automated material model discovery are library-based approaches \citep{flaschel_unsupervised_2021,wang_inference_2021,linka_new_2023}.
Given some data, these methods aim to select a material model from a large library of candidate models using sparse regression.
The core idea of sparse regression is to add a sparsity-promoting $L_1$-regularization term to the loss function that quantifies the mismatch between the model prediction and the data.
By jointly minimizing the model-data mismatch and the regularization term, which is weighted by a regularization parameter $\alpha$, this approach facilitates the discovery of concise material models that fit the data well.
$L_1$-regularization, or more generally $L_p$-regularization, first appeared in the context of model discovery in the early works of \cite{santosa_linear_1986,frank_statistical_1993}.
\cite{tibshirani_regression_1996}, who mathematically analyzed the $L_1$-regularized problem, popularized the method under the name \textit{Least Absolute Shrinkage and Selection Operator} (\LASSO{}).
Since then, $L_1$-regularization has constituted the backbone of automated model discovery, and the concept has been mathematically analyzed and extended in various ways, for example, by \cite{fu_penalized_1998,osborne_new_2000,osborne_lasso_2000,efron_least_2004,daubechies_iterative_2004,zou_regularization_2005,friedman_pathwise_2007,kim_interior-point_2007}.
Originally applied in statistics and data science, $L_1$-regularization found its way into the physical sciences through the seminal work of \cite{brunton_discovering_2016}, who proposed the method SINDy (\textit{Sparse Identification of Nonlinear Dynamics}) to automatically discover short mathematical expressions for dynamic governing equations.
This idea was adopted by the mechanics community to automatically discover material models from data using methods such as EUCLID (\textit{Efficient Unsupervised Constitutive Law Identification and Discovery}) or CANNs (\textit{Constitutive Artificial Neural Networks}), see, for example, \cite{flaschel_unsupervised_2021,wang_inference_2021,wang_establish_2022,linka_new_2023,linka_automated_2023,st_pierre_discovering_2023,st_pierre_principal-stretch-based_2023,flaschel_automated_2023-1,linka_best--class_2024,fuhg_extreme_2024,moon_physics-informed_2025} for hyperelastic materials, \cite{marino_automated_2023} for viscoelastic materials, \cite{flaschel_discovering_2022,meyer_thermodynamically_2023,xu_discovering_2025} for plastic materials, \cite{holthusen_automated_2024} for an application to growth, and \cite{flaschel_automated_2023} for generalized standard materials.

\begin{table}[h!]
\caption{\CHANGE{Existing optimization methods for automated material model discovery based on $L_1$- or $L_p$-regularization.}}
\label{tab:overview_existing}
\centering
\begin{tabular}{|V{4cm}|V{3cm}|V{4cm}|V{4cm}|}
\hline
Method\vphantom{$\frac{\int}{\int}$} & References & Model is linear in the parameters $\bfw$ & Compute regularization path \\ \hline\hline
L-BFGS-B & \cite{wang_inference_2021} & Yes & No \\ \hline
Fixed-point iterations & \cite{flaschel_unsupervised_2021} & Yes & No \\ \hline
Adam & \cite{linka_constitutive_2021} & No & No \\ \hline
Trust-region reflective algorithms & \cite{flaschel_discovering_2022,flaschel_automated_2023} & No & No \\ \hline
\end{tabular}
\end{table}

A distinctive feature of the $L_1$-regularization is its non-smoothness, which gives rise to a non-smooth optimization problem.
While non-smooth optimization problems have been extensively studied within the mechanics community -- particularly in the context of elasto-plasticity, see \cite{kanno_nonsmooth_2011} and more recently \cite{bleyer_applications_2024, bleyer_variational_2024, bleyer_learning_2025} -- they have not yet been rigorously examined in the context of material model discovery.
Although $L_1$-regularization is a well-established tool in the field of material model discovery, existing optimization strategies, such as fixed-point iterations \citep{flaschel_unsupervised_2021}, trust-region reflective algorithms \citep{flaschel_discovering_2022}, or gradient-based methods like the Adam optimizer \citep{linka_new_2023} or L-BFGS-B \citep{wang_inference_2021}, often do not fully exploit the structure of the underlying non-smooth optimization problem.
\CHANGE{
\cref{tab:overview_existing} summarizes commonly used algorithms for mechanical material model discovery. Most of these methods do not explicitly leverage the existing body of knowledge in non-smooth optimization and therefore do not benefit from the theoretical results that have already been established in the non-smooth optimization community.
}
In this work, we address this gap by investigating solvers that are specifically tailored for non-smooth $L_1$-regularized optimization.
This allows for a more principled and potentially more efficient approach to material model discovery, grounded in the theory of non-smooth optimization.
\CHANGE{We describe all algorithms, particularly the LARS algorithm and its internal mechanisms, in a level of detail specifically tailored to the mechanics community.}
For example, we discuss the \textit{Coordinate Descent} (CD) or shooting algorithm \citep{fu_penalized_1998,friedman_pathwise_2007,hastie_elements_2009} and the \textit{Iterative Soft-Thresholding Algorithm} (ISTA) \citep{parikh_proximal_2013,beck_first-order_2017}, which solve the $L_1$-regularized problem for a given value of the regularization parameter $\alpha$ for material model libraries that are linear or nonlinear in the material parameters, respectively.
\CHANGE{
In our numerical examples, we find that the CD algorithm is computationally faster than ISTA. However, ISTA is more generally applicable, as it imposes fewer constraints on the functional form of the objective function.
}

\CHANGE{
Importantly, existing methods for mechanical material model discovery focus on a given value of the regularization parameter, see \cref{tab:overview_existing}.
Here, we do not focus our attention solely on algorithms that solve the $L_1$-regularized problem for a given value of $\alpha$.
Instead, we investigate algorithms for computing the entire regularization path, that is, the solutions of the $L_1$-regularized problem for all possible values of the regularization parameter.
Algorithms for computing the entire regularization path have not been studied in the context of mechanical material modeling before.
One exception is the work of \cite{urreaquintero_automated_2026}, which investigates LARS and related algorithms and was published shortly after we released our preprint.
}
For material model libraries that depend linearly on the parameters, we discuss a modified version of \textit{Least Angle Regression} \LARS{} \citep{osborne_new_2000,efron_least_2004}, which efficiently identifies the critical values of $\alpha$ at which changes in the material model are observed.
Furthermore, inspired by the work of \cite{friedman_pathwise_2007,friedman_regularization_2010,yang_fast_2024-1,yang_fast_2024}, we propose a pathwise extension of ISTA to compute the regularization path for material model libraries with nonlinear parameter dependencies.
\cref{tab:overview} summarizes the key solvers discussed in this work and highlights their specific use cases.
\CHANGE{
For overviews of sparsity-promoting computational algorithms, we refer to the textbooks by \cite{hastie_elements_2009,james_introduction_2023} and the review articles by \cite{hesterberg_least_2008,fan_selective_2010,vidaurre_survey_2013}. For related methods in the context of dimensionality reduction, we refer to \cite{zou_selective_2018}. Finally, for an overview in the context of polynomial chaos expansions, we refer to \cite{luthen_sparse_2021}.
}

\begin{table}[h!]
\caption{Overview of non-smooth optimization methods for automated material model discovery discussed in this contribution.}
\label{tab:overview}
\centering
\begin{tabular}{|V{5cm}|V{5cm}|V{5cm}|}
\hline
\multicolumn{1}{|l|}{\vphantom{$\frac{\int}{\int}$}} & \multicolumn{1}{c|}{Model is linear in parameters $\bfw$} & \multicolumn{1}{c|}{Model is nonlinear in parameters $\bfw$} \\ \hline
Given regularization parameter $\alpha$\vphantom{$\frac{\int}{\int}$} & \textit{Coordinate Descent} CD \citep{fu_penalized_1998} & \textit{Iterative Soft-Thresholding Algorithm} ISTA \citep{parikh_proximal_2013,beck_first-order_2017} \\ \hline
Compute regularization path\vphantom{$\frac{\int}{\int}$} & \textit{Least Angle Regression} \LARSLASSO{}{} \citep{osborne_new_2000,efron_least_2004} & Pathwise ISTA (inspired by \cite{friedman_pathwise_2007}) \\ \hline
\end{tabular}
\end{table}

We note that -- although being arguably the most prominent approach -- $L_1$-regularized regression does not constitute the only method for automated material model discovery.
Alternative approaches include, for example, symbolic regression based on genetic algorithms \citep{koza_genetic_1994,searson_gptipsopen_2010,dubcakova_eureqa_2011,udrescu_ai_2020}.
These approaches have been used by \cite{schmidt_distilling_2009} to discover system dynamics, and by \cite{schoenauer_evolutionary_1996,ratle_grammar-guided_2001,versino_data_2017,kabliman_application_2021,park_multiscale_2021,bomarito_development_2021,abdusalamov_automatic_2023,hou_automated_2024,bahmani_physicsconstrained_2024} in the context of material modeling.
Finally, neural-network-based approaches can be modified to yield interpretable expressions for the material model, as shown by \cite{fuhg_extreme_2024,thakolkaran_can_2025,abdolazizi_constitutive_2025}.

Our paper is structured as follows.
In \cref{sec:mathematical_problems}, we present four mathematical problems that occur in the context of automated material model discovery.
Subsequently, in \cref{sec:solution_algorithms}, we discuss different solvers to approach the four mathematical problems.
In \cref{sec:material_model_discovery}, we place the mathematical problems in the context of material model discovery, and in \cref{sec:results}, we apply the discussed methods to a set of benchmark problems.

\section{Mathematical problems}
\label{sec:mathematical_problems}
The overarching goal of model discovery is to find mathematical models encoded in short mathematical expressions that are capable of describing a given dataset.
Library-based approaches are one of the most popular strategies for model discovery \citep{brunton_discovering_2016,flaschel_unsupervised_2021,linka_new_2023}.
The idea is to construct a general parametric ansatz for the model.
This is also called the model library or model catalog, and it depends on a large number of parameters $\bfw \in \Rset^m$ with $m \gg 1$.
The ability of the model to describe the given dataset is quantified by defining a metric $f(\bfw)$, which measures the mismatch between the model prediction and the data.
The specific form of $f(\bfw)$ depends on the application and the availability of data.
We will discuss different examples throughout this paper.
Our primary objective is to find parameters $\bfw$ that minimize the model-data-mismatch
\be
\label{eq:min_model_data_mismatch}
\bfw^* = \argmin_{\bfw} f(\bfw).
\ee
Due to the large number of parameters in the model library, it is capable of describing complex relationships in a given dataset.
Thus, solving the above problem is likely to discover model parameters $\bfw^*$ for which the model describes the data well.
However, describing the data well is not the only objective of model discovery.
Our second objective is that the model is expressed by a concise mathematical formula, i.e., we seek to find models with a small number of nonzero parameters.
The number of nonzero parameters in $\bfw$ is quantified by the $L_0$-pseudo-norm
\be
\|\bfw\|_0 = \sum_i I(w_i) \quad \text{with} \quad I(w_i) =
\begin{cases}
1 & \text{if } w_i = 0\\
0 & \text{if } w_i \neq 0\\
\end{cases},
\ee
which is not a proper norm because it violates the absolute homogeneity property.
To limit the number of nonzero parameters, we can regularize the minimization problem in \cref{eq:min_model_data_mismatch} by adding $\alpha \, \|\bfw\|_0$ with $\alpha \geq 0$ to the objective function
\be
\label{eq:min_model_data_mismatch_plus_L0}
\bfw^* = \argmin_{\bfw} f(\bfw) + \alpha \, \|\bfw\|_0.
\ee
This regularization term penalizes solutions with many nonzero parameters and thus promotes sparsity in the solution vector.
However, due to the discontinuous and non-convex regularization term, the problem in \cref{eq:min_model_data_mismatch_plus_L0} is computationally expensive to solve for large numbers of parameters.
In practice, the $L_0$-pseudo-norm is often approximated by the $p$-th power of the $L_p$-pseudo-norm with $p>0$ \citep{frank_statistical_1993,flaschel_unsupervised_2021,mcculloch_sparse_2024}
\be
\|\bfw\|^p_p = \sum_i |w_i|^p.
\ee
The $p$-th power of the $L_p$-pseudo-norm is continuous and converges to the $L_0$-pseudo-norm as $p$ approaches zero, sharing the sparsity-promoting property of $L_0$-regularization.
The influence of the choice of $p$ has been studied by \cite{mcculloch_sparse_2024}.
In this work, we focus on the most common choice of $p=1$ \citep{tibshirani_regression_1996,brunton_discovering_2016}, which is the smallest value for which the $L_p$-pseudo-norm is convex and becomes a proper norm, thereby making the minimization problem easier to solve.
We reformulate the problem in \cref{eq:min_model_data_mismatch_plus_L0} as
\be
\label{eq:min_model_data_mismatch_plus_L1}
\bfw^* = \argmin_{\bfw} f(\bfw) + \alpha \|\bfw\|_1 \quad \text{with} \quad \|\bfw\|_1 = \sum_i |w_i|.
\ee
The $L_1$-regularization term was mathematically studied and popularized by \cite{tibshirani_regression_1996,efron_least_2004,hastie_elements_2009,james_introduction_2023}, but it also appears in the early work by \cite{santosa_linear_1986,frank_statistical_1993}.
Due to the continuity and convexity of the $L_1$-norm, the $L_1$-regularized problem in \cref{eq:min_model_data_mismatch_plus_L1} allows for an easier numerical treatment than the $L_0$-regularized problem in \cref{eq:min_model_data_mismatch_plus_L0}, while retaining the sparsity-promoting property, as we will show at several occasions throughout this work.
Similar to the general $L_p$-regularized problem, the $L_1$-regularized problem yields solutions that range from fully dense at small values of $\alpha$ to completely sparse at larger values of $\alpha$.

The problem posed in \cref{eq:min_model_data_mismatch_plus_L1} raises two fundamental questions:
First, noticing that the regularization term is non-smooth, how can the problem be solved efficiently and robustly using methods from the field of non-smooth optimization?
And second, are there optimal strategies for choosing the regularization parameter $\alpha$?
To answer these questions, we distinguish between four types of mathematical problems, which require different numerical solution strategies.
Subsequently, we describe these problems and provide examples from the field of material model discovery.

\subsection{Problem 1}
\label{sec:mathematical_problems_1}

In many practical applications, we are interested in models that depend linearly or affinely on the parameters $\bfw\in\Rset^m$ \citep{frank_statistical_1993,tibshirani_regression_1996,brunton_discovering_2016,flaschel_unsupervised_2021,marino_automated_2023}.
In the case of material modeling, models that depend linearly on the parameters appear, for example, in the constitutive theory of hyperelastic material models \citep{flaschel_unsupervised_2021}, as we will discuss in \cref{sec:material_model_discovery}, or linear viscoelastic material models \citep{marino_automated_2023}. 
For such models, the model prediction $\bfmu\in\Rset^n$ is assumed to be equal to a feature matrix $\bfX$ times the parameters $\bfw$, i.e., $\bfmu = \bfX \bfw$.
We denote the columns of the feature matrix as the feature vectors $\bfX_i$ and observe that the predictions are a linear combination of the feature vectors $\bfmu = \bfX_1 w_1 + \dots + \bfX_m w_m$.
We will assume throughout this work that the feature vectors $\bfX_i$ are linearly independent.

The most common choice for the model-data-mismatch is the sum of squares of the differences between the model prediction and the data
\be
f(\bfw) = \frac{1}{2n} \| \bfy - \bfX \bfw \|_2^2.
\ee
As explained previously, we are interested in minimizing the model-data-mismatch, while at the same time penalizing the $L_1$-norm of the parameters, see \cref{eq:min_model_data_mismatch_plus_L1}.
Consequently, \refp{prob:1} depicts the first type of problem considered in this work, in which we aim to minimize the $L_1$-regularized model-data-mismatch for a given value of $\alpha$.
\refp{prob:1} is also called the \textit{Least Absolute Shrinkage and Selection Operator} (\LASSO{}) \citep{tibshirani_regression_1996}. 

\begin{problem}[label={prob:1}]{}
Given $\bfX \in \Rset^{n \times m}$ with $\|\bfX_i\|_2=1$ and $\bfy \in \Rset^{n}$, define $f(\bfw) = \frac{1}{2n} \| \bfy - \bfX \bfw \|_2^2$.
For a given value of $\alpha \geq 0$, solve
\be
\bfw^* = \argmin_{\bfw} f(\bfw) + \alpha \|\bfw\|_1.
\ee
\end{problem}

We note that in \refp{prob:1}, we assume that the feature vectors are normalized, i.e., $\|\bfX_i\|_2=1$.
This does not affect the generality of the problem.
If, for example, a model with non-normalized feature vectors is given, e.g., $\bfmu=\tilde\bfX\tilde\bfw$, we can normalize the feature vectors according to $\bfX_i = \tilde\bfX_i / \|\tilde\bfX_i\|_2$, while scaling the associated parameters as $\tilde\bfw_i = \bfw_i / \|\tilde\bfX_i\|_2$.
The prediction of the model is not affected because $\bfmu=\tilde\bfX\tilde\bfw=\bfX\bfw$.
Choosing a large regularization parameter $\alpha$ in \refp{prob:1} yields the zero solution $\bfw^*=\boldsymbol{0}$.
As discussed in \cref{sec:boundedness_regularization_parameter}, we are only interested in practically meaningful values of $\alpha$ that yield non-vanishing solutions.
Finally, note that the regularized problem in \refp{prob:1} is mathematically equivalent to the constrained problem 
\be
\label{eq:min_model_data_mismatch_constraint_L1}
\bfw^* = \argmin_{\bfw} f(\bfw) \quad \text{s.t.} \quad \|\bfw\|_1 \leq t,
\ee
where there is a one-to-one relationship between $\alpha > 0$ in \refp{prob:1} and $t > 0$ in \cref{eq:min_model_data_mismatch_constraint_L1} \citep{tibshirani_regression_1996,hastie_elements_2009,james_introduction_2023}.

\subsection{Problem 2}
\label{sec:problem_2}
\begin{figure}[ht!]
\centering
\begin{subfigure}{0.45\textwidth}
\includegraphics[width=\linewidth]{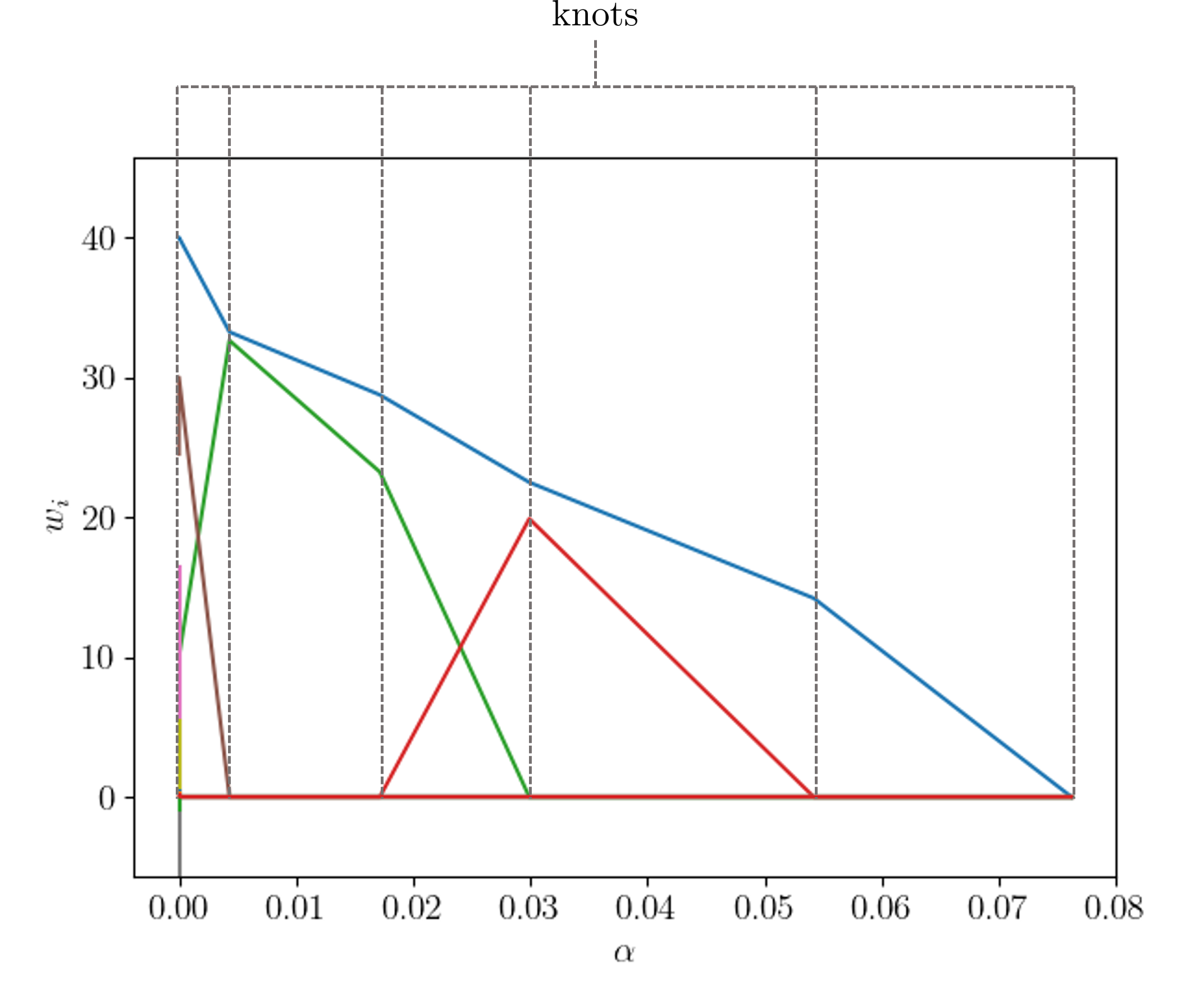}
\caption{Parameters $\bfw$.}
\label{fig:regularization_path_knots}
\end{subfigure}%
\hspace{1.5cm}
\begin{subfigure}{0.45\textwidth}
\includegraphics[width=\linewidth]{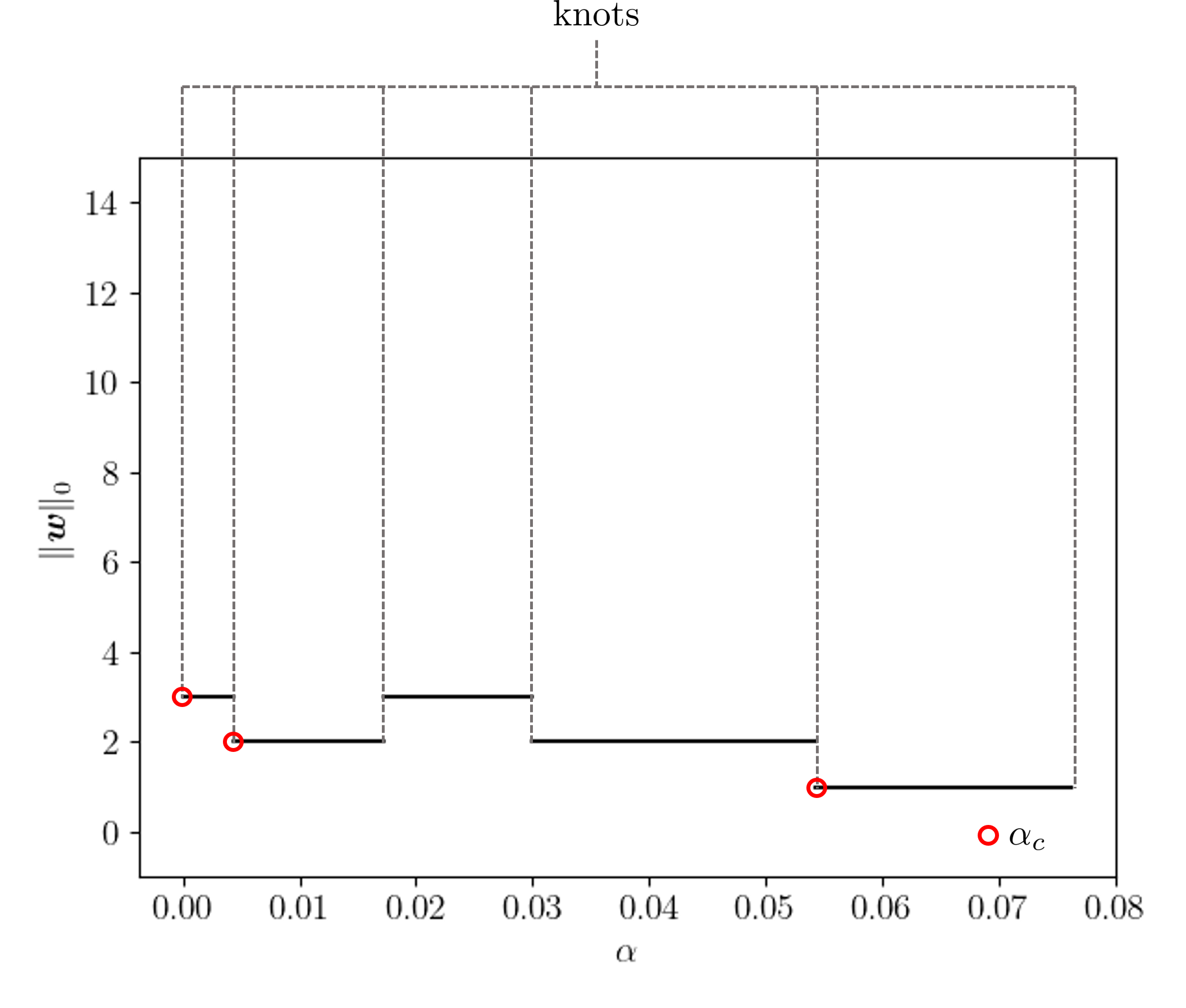}
\caption{Number of nonzero parameters $\|\bfw\|_0$.}
\label{fig:regularization_path_knots_L0}
\end{subfigure}%
\vspace{0.0cm}
\caption{Qualitative regularization path of \refp{prob:1} for a representative dataset. At the knots, the slope of the piecewise linear regularization path changes. The critical values $\alpha_c$, marked as red circles, are the lowest values of $\alpha$ required to achieve a specified sparsity.
}
\label{fig:regularization_path}
\end{figure}

As can be seen from \refp{prob:1}, the solution of the $L_1$-regularized regression problem is dependent on the regularization parameter $\alpha$.
Thus, we may write the solution of the parameters as a function of the regularization parameter $\bfw^*(\alpha)$.
\cref{fig:regularization_path_knots} shows solutions to \refp{prob:1} for different values of $\alpha$ for an exemplary dataset with five features.
The dependence of the parameters on the choice of the regularization parameter is also referred to as the regularization path, or \LASSO{} path for \refp{prob:1} specifically.
Because $\bfX$ and $\bfy$ typically originate from noisy measurements, it is likely that all parameters are nonzero if the problem is not regularized.
This means that $\|\bfw^*(\alpha)\|_0 = m$ if $\alpha = 0$.
Upon increasing the regularization parameter, more and more parameters are forced to be zero by the regularization term.
Interestingly, the regularization path of \refp{prob:1} is piecewise linear \citep{efron_least_2004,kim_interior-point_2007}.
And we observe that, for certain values of the regularization parameter, the slopes of the functions $w^*_i(\alpha)$ change, see \cref{fig:regularization_path}.
We will refer to these values as the knots of the regularization path.
The knots are the only points on the regularization path at which the sparsity of the solution changes.


In general, the sparsity of the solution $\|\bfw^*(\alpha)\|_0$ is not monotone in $\alpha$, see \cref{fig:regularization_path_knots_L0}.
However, we can define a subset of knots $\alpha_c$ with $c=0,\dots,m$ and $0 = \alpha_{m} < \alpha_{m-1} < ... < \alpha_1 < \alpha_0$ such that $\|\bfw^*(\alpha_c)\|_0 = c$ and $\|\bfw^*(\alpha)\|_0 > c$ for all $\alpha < \alpha_c$.
We will refer to these knots as the critical values of the regularization parameter.
When discovering models, we are mainly interested in the critical values of the regularization parameter.
Solutions between critical values are not of interest, because between the critical values $\alpha_c$ and $\alpha_{c+1}$, there exists no solution that at the same time has the same sparsity as the solution $\bfw^*(\alpha_c)$ and exhibits a lower model-data-mismatch than $\bfw^*(\alpha_c)$.
This naturally leads to the question of whether we can directly determine the critical values $\alpha_c$ of the regularization parameter and the associated solution, as depicted in \refp{prob:2}.

\begin{problem}[label={prob:2}]{}
Given $\bfX \in \Rset^{n \times m}$ with $\|\bfX_i\|_2=1$ and $\bfy \in \Rset^{n}$, define $f(\bfw) = \frac{1}{2n} \| \bfy - \bfX \bfw \|_2^2$.
Given the problem
\be
\bfw^*(\alpha) = \argmin_{\bfw} f(\bfw) + \alpha \|\bfw\|_1,
\ee
find the critical values $\alpha_c$ with $c=0,\dots,m$ and $0 = \alpha_{m} < \alpha_{m-1} < ... < \alpha_1 < \alpha_0$
such that $\|\bfw^*(\alpha_c)\|_0 = c$
and $\|\bfw^*(\alpha)\|_0 > c$ for all $\alpha < \alpha_c$.\footnote{For simplicity, we assume the existence of the sequence of values $\alpha_c$.}
\end{problem}

We note that, in principle, it is possible that, for one critical value, two parameters become zero simultaneously.
In other words, there may exist an $\alpha_c$ such that $\|\bfw^*(\alpha_c)\|_0 = c$ and $\|\bfw^*(\alpha)\|_0 \geq c + 2$ for all $\alpha < \alpha_c$.
This is very unlikely for linearly independent feature vectors that originate from noisy measurements.
In the following, we thus neglect this special case.
Finally, we note that, in practice, the goal is typically not to identify all critical values as described in \refp{prob:2}, but rather to focus on the first few critical values that correspond to sparse solutions.


\subsection{Problem 3}

In \refp{prob:1} and \refp{prob:2}, we assumed models that depend linearly on the parameters $\bfw$.
In general, however, models may depend nonlinearly on their parameters.
In the context of material modeling, this is, for example, the case for certain hyperelastic material models such as power-type Ogden models \citep{flaschel_automated_2023-1}, exponential-type models \citep{linka_new_2023}, or dissipative material models \citep{flaschel_discovering_2022,flaschel_automated_2023}.
For such models, the model prediction $\bfmu(\bfw)$ is a nonlinear function of the parameters.
The model-data-mismatch is therefore not quadratic and may even be non-convex.
As the third type of problem that we consider in this work, see \refp{prob:3}, we consider the generalization of \refp{prob:1}, i.e., an $L_1$-regularized problem in which the model-data-mismatch is not quadratic.
We restrict our attention to models for which the model prediction $\bfmu(\bfw)$ is differentiable with respect to the parameters such that also $f(\bfw)$ is differentiable.

\begin{problem}[label={prob:3}]{}
Given a differentiable function $f(\bfw)$.
For a given value of $\alpha \geq 0$, solve
\be
\bfw^* = \argmin_{\bfw} f(\bfw) + \alpha \|\bfw\|_1.
\ee
\end{problem}

\subsection{Problem 4}
\refp{prob:3} requires an a priori selection of the regularization parameter $\alpha$.
In practice, however, it is often not clear beforehand which value of $\alpha$ results in a model that is both sparse and demonstrates high fitting accuracy.
Consequently, it can be beneficial to compute the regularization path for \refp{prob:3}.
Unlike the regularization path visualized in \cref{fig:regularization_path}, for models that depend nonlinearly on the parameters, the regularization path is not necessarily piecewise linear.
For these models, identifying the critical values of $\alpha$ is not straightforward.
A practical remedy is to solve \refp{prob:3} for a predefined set of values of $\alpha$, as depicted in \refp{prob:4}.
The values of $\alpha$ should be within a meaningful range, i.e., between zero and the smallest value of $\alpha$, denoted by $\alpha^{(0)}$, that yields the zero parameter vector, the computation of which is discussed in \cref{sec:boundedness_regularization_parameter}.
As we discuss below, the computational burden of \refp{prob:4} is not simply $n_{\alpha}$ times the cost of \refp{prob:3}, since solutions at previous values of $\alpha$ can be used as initial estimates for subsequent computations, see \cite{friedman_pathwise_2007,friedman_regularization_2010,yang_fast_2024-1,yang_fast_2024}.

\begin{problem}[label={prob:4}]{}
Given a differentiable function $f(\bfw)$, solve
\be
\bfw^* = \argmin_{\bfw} f(\bfw) + \alpha^{(k)} \|\bfw\|_1,
\ee
for a predefined set of values $\alpha^{(k)}$ with $k=0,\dots,n_{\alpha} - 1$ and $\alpha^{(k)} \in (0,\alpha^{(0)}]$, where $\alpha^{(0)} = \max_i \left|\frac{\partial f}{\partial w_i}(\boldsymbol{0})\right|$.
\end{problem}

\section{Solution algorithms}
\label{sec:solution_algorithms}
\subsection{\textit{Coordinate Descent} (CD)}

A popular algorithm for solving \refp{prob:1} is the \textit{coordinate descent} (CD) algorithm, also known as the shooting algorithm.
It has been used and mathematically analyzed by \cite{fu_penalized_1998,friedman_pathwise_2007}.

We start with an initial guess for the parameters $\bfw^{(0)}$, for which we typically consider the ordinary least squares solution $\bfw^{(0)} = [\bfX^T\bfX]^{-1}\bfX^T\bfy$, and iteratively compute a sequence $\bfw^{(k)}$ with $k=0,1,\dots$ that converges to the solution of \refp{prob:1}.
The basic idea of the CD algorithm is to loop over all parameters at each step and treat all parameters except one as constants while minimizing the objective of \refp{prob:1}.
Specifically, at each step $k$, we first set $\bfw^{(k)}$ to be equal to $\bfw^{(k-1)}$, and then for all $l=1,\dots,m$ set $w^{(k)}_l$ to be equal to
\be
\label{eq:CD_step_general}
w^{*(k)}_l = \argmin_{w^{(k)}_l} \frac{1}{2n} \| \bfX \bfw^{(k)} - \bfy \|_2^2 + \alpha \|\bfw^{(k)}\|_1.
\ee
Geometrically, each solving of \cref{eq:CD_step_general} can be interpreted as minimizing along one coordinate $w^{(k)}_l$, while treating all other parameters as constants, which explains the name of the \textit{coordinate descent} algorithm.

The strength of CD is that the minimization problem in \cref{eq:CD_step_general} is a convex and non-smooth minimization problem that admits a closed-form and computationally efficient solution.
In the following, we recapitulate the basic concepts of convex and non-smooth optimization and detail the solution of  \cref{eq:CD_step_general}.
First, we denote the objective function of \cref{eq:CD_step_general} by $L(w^{(k)}_l)$ and use index notation to rewrite it as
\be
\begin{aligned}
L(w^{(k)}_l)
&= \frac{1}{2n} \sum_i \left[ \sum_j X_{ij} w^{(k)}_j - y_i \right]^2 + \alpha \sum_j |w^{(k)}_j| \\
&=
\underbrace{\frac{1}{2n} \sum_i \left[ \sum_{j \neq l} X_{ij} w^{(k)}_j + X_{il} w^{(k)}_l - y_i \right]^2}_{L_f(w^{(k)}_l)}
+ \underbrace{\alpha \sum_{j \neq l} |w^{(k)}_j| + \alpha |w^{(k)}_l|}_{L_\alpha(w^{(k)}_l)}, \\
\end{aligned}
\ee
where $w^{(k)}_j$ with $j \neq l$ are treated as constants. The first part of the objective function $L_f(w^{(k)}_l)$ is smooth and differentiable. Its derivative computes to
\be
\frac{\partial L_f}{\partial w^{(k)}_l}
= \frac{1}{n} \sum_i X_{il} \left[ \sum_{j \neq l} X_{ij} w^{(k)}_j + X_{il} w^{(k)}_l - y_i \right]
= S^{(k)}_l + \frac{1}{n} \|\bfX_l\|^2_2 w^{(k)}_l,
\ee
where $S^{(k)}_l = \frac{1}{n} \sum_i X_{il} \left[ \sum_{j \neq l} X_{ij} w^{(k)}_j - y_i \right]$.
The second part of the objective function $L_\alpha(w^{(k)}_l)$, however, is non-smooth and non-differentiable at $w^{(k)}_l = 0$, which consequently means that $L(w^{(k)}_l)$ is non-smooth and non-differentiable.

From the theory of convex and non-smooth optimization \citep{rockafellar_convex_1970,boyd_convex_2004}, we recall the concepts of subderivatives and subdifferentials.
A real number $d$ is a subderivative of $L(w^{(k)}_l)$ at the point $\hat w^{(k)}_l$ if $L(w^{(k)}_l) - L(\hat w^{(k)}_l) \geq d [w^{(k)}_l - \hat w^{(k)}_l]$ for all $w^{(k)}_l$.
This can be interpreted as drawing a line of slope $d$ through the point $\{ \hat w^{(k)}_l,L(\hat w^{(k)}_l) \}$.
If the line is smaller than or equal to the graph of $L(w^{(k)}_l)$ for all $w^{(k)}_l$, then $d$ is a subderivative.
The set of all subderivatives of $L(w^{(k)}_l)$ is called the subdifferential and will in the following be denoted by $\partial_{w^{(k)}_l} L$.
A necessary and sufficient condition for a minimum of the convex function $L(w^{(k)}_l)$ is that the subdifferential contains zero, $0 \in \partial_{w^{(k)}_l} L$, i.e., $d=0$ is a subderivative at the minimum.

To derive the subdifferential of $L(w^{(k)}_l)$, we consider $L_f(w^{(k)}_l)$ and $L_\alpha(w^{(k)}_l)$ separately.
The first term,
$L_f(w^{(k)}_l)$, is smooth and differentiable, 
and its subderivative is uniquely determined by the derivative ${\partial L_f}/{\partial w^{(k)}_l}$ at each point.
Thus, its subdifferential is the singleton $\partial_{w^{(k)}_l} L_f = \{{\partial L_f}/{\partial w^{(k)}_l}\}$.
The second term, 
$L_\alpha(w^{(k)}_l)$, is non-smooth and non-differentiable. 
Its subdifferential is identified as
\be
\partial_{w^{(k)}_l} L_\alpha = 
\begin{cases}
    \{-\alpha\} & \text{if } w^{(k)}_l < 0 \\
    [-\alpha,\alpha] & \text{if } w^{(k)}_l = 0 \\
    \{\alpha\} & \text{if } w^{(k)}_l > 0
\end{cases}.
\ee
Finally, the subdifferential $\partial_{w^{(k)}_l} L$ is the sum of the subdifferentials $\partial_{w^{(k)}_l} L_f$ and $\partial_{w^{(k)}_l} L_\alpha$,
\be
\partial_{w^{(k)}_l} L = 
\begin{cases}
    \{S^{(k)}_l + \frac{1}{n} \|\bfX_l\|^2_2 w^{(k)}_l - \alpha \} 
    & \text{if } w^{(k)}_l < 0 \\
    [S^{(k)}_l - \alpha, S^{(k)}_l + \alpha ]
    & \text{if } w^{(k)}_l = 0 \\
    \{S^{(k)}_l + \frac{1}{n} \|\bfX_l\|^2_2 w^{(k)}_l + \alpha \}
    & \text{if } w^{(k)}_l > 0 
\end{cases}.
\ee
To solve the minimization problem in \cref{eq:CD_step_general}, we seek to find $w^{*(k)}_l$ such that $0 \in \partial_{w^{(k)}_l} L$. To this end, we consider the three cases:
\begin{itemize}
    \item For $w^{*(k)}_l<0$,
    it must be $S^{(k)}_l + \frac{1}{n} \|\bfX_l\|^2_2 w^{*(k)}_l - \alpha = 0$, from which we deduce
    $
    w^{*(k)}_l = n \, (\alpha - S^{(k)}_l)/ \|\bfX_l\|^2_2
    $.
    This can only be an admissible solution if $w^{*(k)}_l<0$ and thus if $S^{(k)}_l > \alpha$.
    \item For $w^{*(k)}_l = 0$,
    it must be $0 \in [S^{(k)}_l - \alpha, S^{(k)}_l + \alpha]$. This is only possible if $S^{(k)}_l - \alpha \leq 0$ and $S^{(k)}_l + \alpha \geq 0$, and thus if $- \alpha \leq S^{(k)}_l \leq \alpha$.
    \item For, $w^{*(k)}_l > 0$,
    analogously to the first case, we obtain
    $
    w^{*(k)}_l = n \, (-\alpha - S^{(k)}_l)/\|\bfX_l\|^2_2.
    $
    This can only be an admissible solution if $w^{*(k)}_l>0$ and thus if $S^{(k)}_l < -\alpha$.
\end{itemize}
Summarizing all three cases, we deduce the closed-form solution to \cref{eq:CD_step_general}
\be
\label{eq:CD_step_closed_form}
w^{*(k)}_l = 
\begin{cases}
    n\, \frac{\alpha - S^{(k)}_l}{ \|\bfX_l\|^2_2} & \text{if } S^{(k)}_l > \alpha \\
    0 & \text{if } - \alpha \leq S^{(k)}_l \leq \alpha \\
    n\, \frac{-\alpha - S^{(k)}_l}{ \|\bfX_l\|^2_2} & \text{if } S^{(k)}_l < -\alpha \\
\end{cases}.
\ee
By introducing the so-called soft-thresholding function $\text{soft}_\alpha(x) = \sign(x)\max \{ |x| - \alpha, 0 \}$, the closed-form solution is concisely written as
\be
\label{eq:CD_step_closed_form2}
w^{*(k)}_l = - \frac{\text{soft}_\alpha(S^{(k)}_l)}{\frac{1}{n} \|\bfX_l\|^2_2}.
\ee
We observe that $w^{*(k)}_l$ is set exactly to zero if $- \alpha \leq S^{(k)}_l \leq \alpha$.
Increasing $\alpha > 0$ increases the chance that $- \alpha \leq S^{(k)}_l \leq \alpha$ is fulfilled, and thus increases the sparsity of parameters.
We finally notice that the closed-form solution further simplifies due to the normalization of $\bfX$, i.e., $\|\bfX_l\|^2_2 = 1$.
The CD algorithm is summarized in \cref{alg:CD}.

\begin{algorithm}
\caption{\textit{Coordinate Descent} (CD)}\label{alg:CD}
\begin{algorithmic}
\State Given $\bfX$ and $\bfy$
\State Set initial guess $\bfw^{(0)}=[\bfX^T\bfX]^{-1}\bfX^T\bfy$
\State Choose the maximum number of steps NSTEP and convergence tolerance TOL
\For{$k=1,\dots,\text{NSTEP}$}
\State $\bfw^{(k)} \gets \bfw^{(k-1)}$
\For{$l=1,\dots,m$}
\State $S^{(k)}_l = \frac{1}{n} \sum_i X_{il} \left[ \sum_{j \neq l} X_{ij} w^{(k)}_j - y_i \right]$
\State
$w^{(k)}_i \gets
\begin{cases}
    \frac{\alpha - S^{(k)}_l}{\frac{1}{n} \|\bfX_l\|^2_2} & \text{if } S^{(k)}_l > \alpha \\
    0 & \text{if } - \alpha \leq S^{(k)}_l \leq \alpha \\
    \frac{-\alpha - S^{(k)}_l}{\frac{1}{n} \|\bfX_l\|^2_2} & \text{if } S^{(k)}_l < -\alpha \\
\end{cases}
$
\EndFor
\If{$\| \bfw^{(k+1)} - \bfw^{(k)} \|_2 < \text{TOL}$ \textbf{or} $\left| f(\bfw^{(k+1)}) + \alpha\|\bfw^{(k+1)}\|_1 - f(\bfw^{(k)}) - \alpha\|\bfw^{(k)}\|_1 \right| < \text{TOL}$}
\State \textbf{break}
\EndIf
\EndFor
\end{algorithmic}
\end{algorithm}

CD efficiently solves \refp{prob:1} with proven convergence, see \cite{fu_penalized_1998}.
In principle, CD can also be used to numerically approach \refp{prob:2}, by simply solving \refp{prob:1} for a large number of different values for $\alpha$ and determining approximations of the critical values $\alpha_c$ at which parameters are set to zero.
However, such a brute force attempt to solving \refp{prob:2} is computationally infeasible, especially for larger numbers of parameters.
As we will discuss below, there exist more efficient algorithms to approach \refp{prob:2}.
In special cases, CD can also be used to approach \refp{prob:3}.
As proposed by \cite{flaschel_automated_2023-1}, if the model is linearly dependent on many of the parameters and nonlinearly dependent on only a few of the parameters, the latter parameters can be discretized such that they vanish from the set of parameters to be determined in the optimization problem.
Due to the discretization, \refp{prob:3} is transformed to \refp{prob:1} with a larger number of parameters, see \cite{flaschel_automated_2023-1} for details.
However, this strategy is only applicable for a small number of nonlinear parameters.
In the following, we will discuss algorithms beyond CD to efficiently solve \refp{prob:2} and \refp{prob:3}.


Dependent on the choice of the initial guess, CD can be considered as either a \textit{top-down} or \textit{bottom-up} approach.
If the ordinary least squares solution $\bfw^{(0)} = [\bfX^T\bfX]^{-1}\bfX^T\bfy$ is chosen as the initial guess, CD can be interpreted as a \textit{top-down} approach.
That is, starting with the ordinary least squares solution, which in general has many nonzero entries, the algorithm progressively sets more and more components of the parameter vector to zero during its iterations.
If, however, $\bfw^{(0)} = \boldsymbol{0}$ is chosen as the initial guess, CD can be interpreted as a \textit{bottom-up} approach that progressively adds more nonzero components to the parameter vector.

We note that CD can also be used to solve \refp{prob:2} by solving \refp{prob:1} for a predefined set of values $\alpha$.
Previously computed solutions for specific values of $\alpha$ can serve as initial guesses for neighboring values to improve efficiency.
However, this strategy does not exploit the piecewise linear structure of the regularization path, as the \LARS{} algorithm discussed below does.

\subsection{\textit{Least Angle Regression} (\LARS{} and \LARSLASSO{})}

\cite{efron_least_2004} proposed and mathematically analyzed the so-called \textit{Least Angle Regression} (\LARS{}), an algorithm for model selection that shares similarities with the earlier proposed \textit{homotopy method} by \cite{osborne_new_2000}.
This algorithm starts with the zero solution $\bfw^{(0)}=\bf0$ and builds a sequence of parameter vectors $\bfw^{(0)},\bfw^{(1)},\bfw^{(2)},\dots$ by successively considering more nonzero parameters.
In each step of the algorithm, only those parameters whose features show the highest absolute correlation with the residual $\bfr^{(k)} = \bfy - \bfX \bfw^{(k)}$ are modified.
These parameters, which are called the active set, are updated toward the least squares solution of the problem until a new feature not belonging to the active set becomes equally correlated with the residual.
Then, this feature is included in the active set and the process is repeated.
As the algorithm progresses, the number of nonzero parameters increases and the model-data mismatch generally decreases. 
In this sense, \LARS{} can be interpreted as an algorithm that belongs to the family of so-called \textit{forward stepwise selection algorithms}, see \cite{james_introduction_2023}.
Interestingly, \cite{efron_least_2004} show that the solutions $\bfw^{(k)}$ obtained by \LARS{} are similar to the solutions sought in \refp{prob:2}, and the authors propose a simple modification of \LARS{}, which we call \LARSLASSO{}, to efficiently solve \refp{prob:2}.
\LARSLASSO{} finds solutions to the \LASSO{} \refp{prob:1}, but at the same time identifies the critical values of $\alpha$ at which the number of nonzero parameters changes.
Thus, it provides an efficient tool to identify the \LASSO{} regularization path.
In the following, we explain \LARS{} in detail for the special case of two features and the general case of $m$ features, and show how to modify \LARS{} to obtain \LARSLASSO{}.

\subsubsection{\LARS{} considering two features}
\label{sec:LARS_two_features}

Following \cite{efron_least_2004}, we focus on an example with two features and two parameters, i.e., $\bfX \in \Rset^{n\times2}$ and $\bfw \in \Rset^2$, to explain the main ideas behind \LARS{} and to visualize the algorithm.

Initially, all parameters are set to zero $\bfw^{(0)}=\bf0$, and the initial model prediction is computed as $\bfmu^{(0)}=\bfX \bfw^{(0)}=\bf0$.
At step $k$, the residual vector between the data $\bfy$ and the prediction $\bfmu^{(k)}$ is defined as $\bfr^{(k)} = \bfy - \bfX \bfw^{(k)}$.
The data can be additively divided into two parts $\bfy = \bfy_{\parallel} + \bfy_{\perp}$,
such that the first part $\bfy_{\parallel}$ can be expressed as a linear combination of the linearly independent feature vectors $\bfX_1$ and $\bfX_2$, i.e., $\bfy_{\parallel} = \bfX_1 a_1 + \bfX_2 a_2$ with $a_1,a_2\in\Rset$, while the second part $\bfy_{\perp}$ is perpendicular to the feature vectors, i.e., $\bfX_i^T \bfy_{\perp} = 0$.
The first part $\bfy_{\parallel}$ is computed as the ordinary least squares solution $\bfy_{\parallel}=\bfX\left[\bfX^T\bfX\right]{}^{-1}\bfX^T\bfy$, i.e., the projection of $\bfy$ onto the plane spanned by the feature vectors.
Substituting $\bfy = \bfy_{\parallel} + \bfy_{\perp}$ into the residual, we obtain $\bfr^{(k)} = \bfy_{\parallel} - \bfX \bfw^{(k)} + \bfy_{\perp}$, and we define a parallel $\bfr^{(k)}_{\parallel} = \bfy_{\parallel} - \bfX \bfw^{(k)}$ and a perpendicular $\bfr^{(k)}_{\perp} = \bfy_{\perp}$ part of the residual. 
We notice that, no matter how we modify the parameters $\bfw^{(k)}$, the perpendicular part of the residual $\bfr^{(k)}_{\perp}$ remains unchanged.

\begin{figure}[ht!]
\centering
\includegraphics[width=0.9\linewidth]{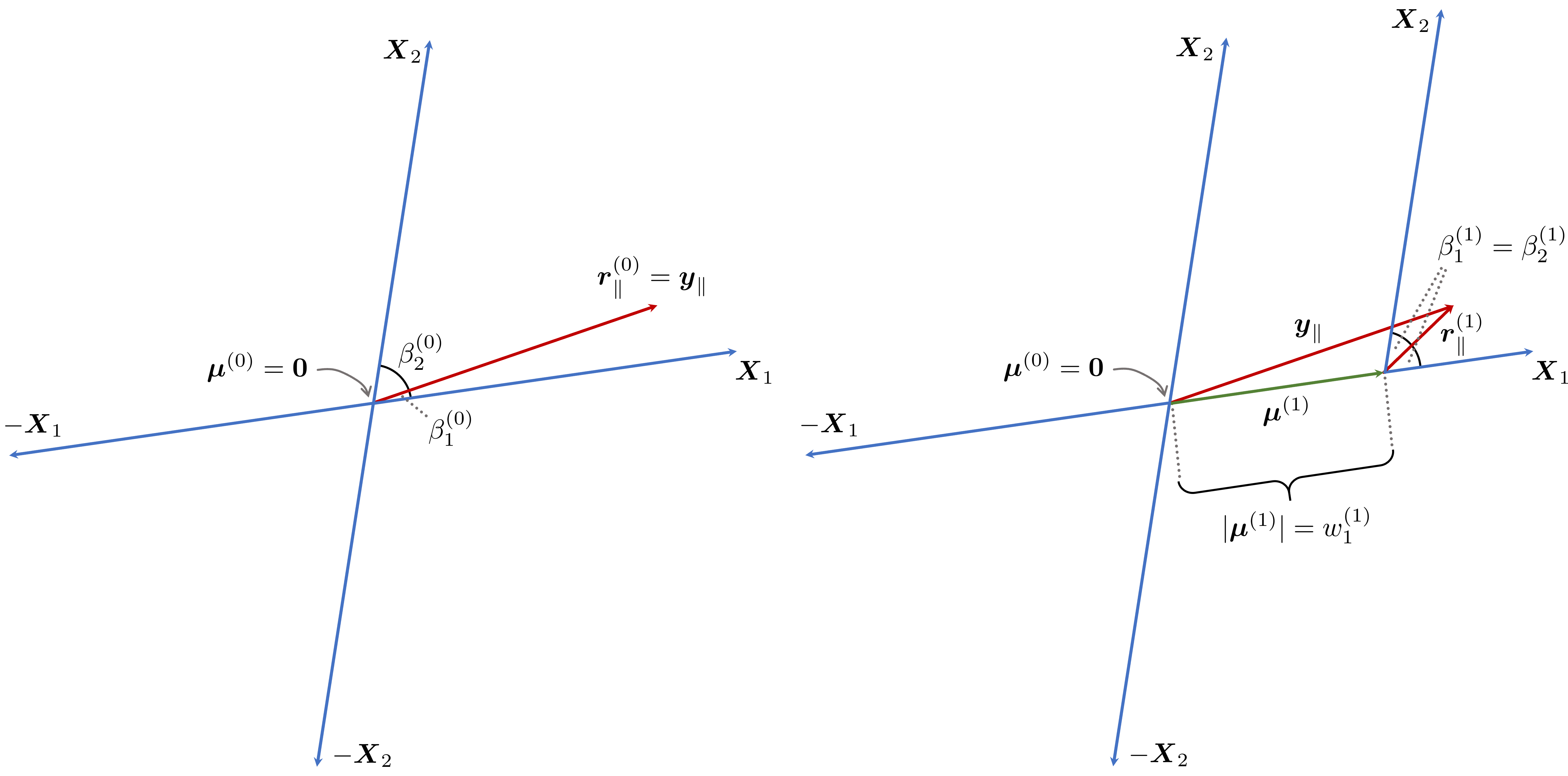}
\caption{Illustration of the first step of \LARS{} considering two features. All vectors are illustrated in the two-dimensional plane spanned by the feature vectors $\bfX_1$ and $\bfX_2$. The left figure illustrates the initial prediction $\bfmu^{(0)}=\bf0$ and the initial residual $\bfr^{(0)}_{\parallel}=\bfy_{\parallel}$. The feature vector $\bfX_1$ shows the least angle to the residual $\bfr^{(0)}_{\parallel}$. Therefore, we assume that, at the next step, the parameter $w^{(1)}_1 \neq 0$ is nonzero, such that the prediction at the next step $\bfmu^{(1)}$ is parallel to $\bfX_1$. As illustrated in the right figure, $w^{(1)}_1$ is chosen such that the new residual $\bfr^{(1)}_{\parallel}$ shows the same angle to both feature vectors.}
\label{fig:LARS_two_features}
\end{figure}

\cref{fig:LARS_two_features} illustrates the first step of \LARS{} for the example with two features.
Recall that all parameters are initially zero $\bfw^{(0)}=\bf0$.
In the first step of \LARS{}, we set one of the parameters to be nonzero such that the prediction changes along the direction of the corresponding feature vector.
To select which of the parameters should be set to nonzero, we seek to identify the feature vector that, upon adjusting its corresponding parameter, promises the greatest reduction of the norm of the residual at the next step $\bfr^{(1)}_{\parallel}$.
To this end, we investigate the angles between the feature vectors $\bfX_i$ and the \mbox{residual $\bfr^{(0)}_{\parallel}$}
\be
\label{eq:angle}
\beta^{(0)}_i = \sphericalangle(\bfX_i,\bfr^{(0)}_{\parallel}) = \cos^{-1} \frac{\bfX_i^T \bfr^{(0)}_{\parallel}}{\|\bfX_i\| \ \|\bfr^{(0)}_{\parallel}\|}.
\ee
Since we also allow for negative parameters, which means that the model prediction can also change along the direction of  negative feature vectors, 
we are also interested in the angles between the negative feature vectors and the residual, i.e., $\beta^{(0)}_{-i} = \sphericalangle(-\bfX_i,\bfr^{(0)}_{\parallel})$.
Considering two features, this amounts to a total of four angles $\beta^{(0)}_1$, $\beta^{(0)}_2$, \mbox{$\beta^{(0)}_{-1}=\pi - \beta^{(0)}_1$}, \mbox{$\beta^{(0)}_{-2}=\pi - \beta^{(0)}_2$}.
The feature vector or negative feature vector with the least angle to the residual promises the greatest reduction of the norm of the residual upon adjusting its corresponding parameter. This is the reason why the method is called \textit{Least Angle Regression}.

It is not necessary to explicitly compute the angles.
Instead, we compute the correlation vector $\bfc^{(0)} = \bfX^T \bfr^{(0)}_{\parallel}$ with $c^{(0)}_i = \bfX_i^T \bfr^{(0)}_{\parallel}$.
Because the feature vectors are normalized $\|\bfX_i\|=1$ and the function $\cos^{-1}(\cdot)$ in \cref{eq:angle} is strictly monotonically decreasing, we can deduce that \mbox{$\beta^{(0)}_i < \beta^{(0)}_j$} if and only if \mbox{$c^{(0)}_i > c^{(0)}_j$}.
Further, it is $c^{(0)}_{-i} = -\bfX_i^T \bfr^{(0)}_{\parallel} = - c^{(0)}_i$.
Thus, to identify the feature vector or negative feature vector corresponding to the smallest angle, it is sufficient to identify the feature vector with the greatest absolute correlation $\max_i |c^{(0)}_i|$, where the sign of the correlation $s^{(0)}_i=\sign(c^{(0)}_i)$ indicates whether the feature vector or its negative corresponds to the smallest angle.

After having identified the feature vector $\bfX_{i^*}$ with the greatest absolute correlation, i.e., $i^* = \argmax_{i \in \{1,2\}} |c^{(0)}_i|$, we adjust the corresponding parameter $w^{(1)}_{i^*} = w^{(0)}_{i^*} + \Delta w^{(0)}$. 
The prediction at the next step is therefore $\bfmu^{(1)}=\bfmu^{(0)}+\bfX_{i^*} \Delta w^{(0)}=\bfX_{i^*} \Delta w^{(0)}$ and the residual is $\bfr^{(1)}_{\parallel} = \bfy_{\parallel} - \bfX_{i^*} \Delta w^{(0)}$.
The adjustment $\Delta w^{(0)}$ could, for example, be chosen such that $\|\bfr^{(1)}_{\parallel}\|^2$ is minimized, which would result in $\Delta w^{(0)}=(\bfX_{i^*}^T \bfX_{i^*})^{-1}\bfX_{i^*}^T \bfy_{\parallel} = c^{(0)}_{i^*}$, where we made use of the fact that the feature vectors are normalized.
Choosing $\Delta w^{(0)}$ in this way is the core idea of so-called \textit{forward stepwise selection methods}, see the overview by \cite{james_introduction_2023}.
However, \textit{forward stepwise selection methods} are known to disregard features from the active set, even when they exhibit a strong correlation with the data.
As a result, \cite{efron_least_2004} describe them as "overly greedy".
On the contrary, the idea of \LARS{} is to adjust $\Delta w^{(0)}$ until another feature vector $\bfX_{j^*}$ with $i^* \neq j^*$ becomes equally correlated in absolute value with the residual.
This means that we seek to choose $\Delta w^{(0)}$ such that $|c^{(1)}_{i^*}|=|c^{(1)}_{j^*}|$.
For the example with two features, this means that either $c^{(1)}_{i^*}=c^{(1)}_{j^*}$ or $c^{(1)}_{i^*}=-c^{(1)}_{j^*}$.
The first condition leads to 
\be
\label{eq:LARS_step_two_features1}
\begin{aligned}
    c^{(1)}_{i^*}=c^{(1)}_{j^*}\quad
    \Rightarrow\quad & \bfX_{i^*}^T \bfr^{(1)}_{\parallel} = \bfX_{j^*}^T \bfr^{(1)}_{\parallel} \\
    \Rightarrow\quad & \bfX_{i^*}^T (\bfy_{\parallel} - \bfX_{i^*} \Delta w^{(0)}) = \bfX_{j^*}^T (\bfy_{\parallel} - \bfX_{i^*} \Delta w^{(0)}) \\
    \Rightarrow\quad & \Delta w^{(0)}
    = \frac{\bfX_{i^*}^T \bfy_{\parallel} - \bfX_{j^*}^T \bfy_{\parallel}}{\bfX_{i^*}^T \bfX_{i^*} - \bfX_{j^*}^T \bfX_{i^*}}
    = \frac{c^{(0)}_{i^*} - c^{(0)}_{j^*}}{1 - \bfX_{j^*}^T \bfX_{i^*}}, \\
\end{aligned}
\ee
while the second condition leads to
\be
\label{eq:LARS_step_two_features2}
\begin{aligned}
    c^{(1)}_{i^*}=-c^{(1)}_{j^*}\quad
    \Rightarrow\quad & \Delta w^{(0)}
    = \frac{\bfX_{i^*}^T \bfy_{\parallel} + \bfX_{j^*}^T \bfy_{\parallel}}{\bfX_{i^*}^T \bfX_{i^*} + \bfX_{j^*}^T \bfX_{i^*}}
    = \frac{c^{(0)}_{i^*} + c^{(0)}_{j^*}}{1 + \bfX_{j^*}^T \bfX_{i^*}}. \\
\end{aligned}
\ee
In \cref{eq:LARS_step_two_features1,eq:LARS_step_two_features2}, we made use of the normalization of the feature vectors $\bfX_{i^*}^T \bfX_{i^*}=1$.
Further, due to the normalization, it is $-1<\bfX_{j^*}^T \bfX_{i^*}<1$.
This means that the denominator in \cref{eq:LARS_step_two_features1,eq:LARS_step_two_features2} is positive.
Because of $|c^{(0)}_{i^*}|>|c^{(0)}_{j^*}|$, the sign of the numerator is $\sign(c^{(0)}_{i^*})$.
Therefore, we can deduce that $\sign(\Delta w^{(0)}) = \sign(c^{(0)}_{i^*})$, which means that the $w_{i^*}$ is updated such that it shares the sign with the correlation of the corresponding feature vector.

Among the two potential values for $\Delta w^{(0)}$ in \cref{eq:LARS_step_two_features1,eq:LARS_step_two_features2}, we choose the one with the smallest absolute value.
After computing $\Delta w^{(0)}$ and thus $\bfw^{(1)}$, which has one nonzero entry, we finally compute the solution vector with all parameters being nonzero as the ordinary least squares solution $\bfw^{(2)}=\left[\bfX^T\bfX\right]^{-1}\bfX^T\bfy$.

\subsubsection{\LARS{} for more than two features}
\label{sec:LARS}

We now consider \LARS{} \citep{efron_least_2004} for the general case with more than two features, i.e., $\bfX \in \Rset^{n \times m}$ and $\bfw \in \Rset^m$.  
As for the case with two features, \LARS{} starts by setting all parameters to zero $\bfw^{(0)}=\bf0$ and builds a sequence of parameter vectors $\bfw^{(0)},\bfw^{(1)},\dots$ by successively considering more nonzero parameters.
From \cref{sec:LARS_two_features}, we recapitulate the projection of $\bfy$ onto the plane spanned by the feature vectors $\bfy_{\parallel}=\bfX\left[\bfX^T\bfX\right]^{-1}\bfX^T\bfy$,
the prediction $\bfmu^{(k)} = \bfX \bfw^{(k)}$,
the parallel part of the residual $\bfr^{(k)}_{\parallel} = \bfy_{\parallel} - \bfmu^{(k)}$,
and the correlations between each feature and the residual $\bfc^{(k)} = \bfX^T \bfr^{(k)}_{\parallel}$ with $c^{(k)}_i = \bfX_i^T \bfr^{(k)}_{\parallel}$.

At each step of \LARS{}, we identify those feature vectors that exhibit the greatest absolute correlation with the residuals.
Specifically, we define the active set and its complement
\be
\mathcal{A}^{(k)} = \left\{ i^*\in \{1,\dots,m\} ~ \big| ~ |c^{(k)}_{i^*}| = \bar{c}^{(k)}_{\text{max}} = \max_i |c^{(k)}_i| \right\},
\quad \mathcal{A}^{\complement(k)} = \left\{ j^*\in \{1,\dots,m\} ~ \big| ~ j^* \notin \mathcal{A}^{(k)} \right\}.
\ee
In the following, we will use the indices $i^*$ and $j^*$ to refer to elements in the active set and its complement at the current step $k$, respectively.

The signs of the correlations indicate whether the corresponding feature vector or its negative counterpart exhibits a greater correlation with the residual.
Thus, the signs indicate whether the parameters corresponding to the feature vectors should increase or decrease in order to reduce the residual at the next step.
We define the vector $\bfs^{(k)}$ such that $s^{(k)}_i=\sign(c^{(k)}_i)$,
and we flip the signs of the feature vectors $\bar{\bfX}^{(k)}_i = s^{(k)}_i \bfX_i$,
such that their correlations with the residual are positive $\bar{c}^{(k)}_i = \bar{\bfX}^{T(k)}_i \bfr^{(k)}_{\parallel} = s^{(k)}_i c^{(k)}_i = |c^{(k)}_i|$.

\LARS{} updates the prediction at each step according to the rule
\be
\label{eq:LARS_step_prediction}
\bfmu^{(k+1)}=\bfmu^{(k)} + \gamma^{(k)} \bfu^{(k)},
\ee
where the unit vector $\bfu^{(k)}$ with $\|\bfu^{(k)}\|=1$ defines the direction of the update and $\gamma^{(k)} > 0$ defines the step size.
This consequently means that the correlations are updated according to
\be
\label{eq:LARS_step_correlation}
c^{(k+1)}_i = \bfX_{i}^T \bfr^{(k+1)}_{\parallel} = \bfX_{i}^T (\bfy_{\parallel} - \bfmu^{(k)} - \gamma^{(k)} \bfu^{(k)}) = c^{(k)}_{i} - \gamma^{(k)} a^{(k)}_{i},
\ee
where we defined $a^{(k)}_{i}=\bfX_{i}^T \bfu^{(k)}$.
In the following, we detail the choice of $\bfu^{(k)}$ and $\gamma^{(k)}$.

At the current step, all active feature vectors $\bfX^{(k)}_{i^*}$ exhibit the same absolute correlation with the residual,
i.e., $\bar{c}^{(k)}_{i^*}$ is the same for all $i^*\in\mathcal{A}^{(k)}$.
The idea of \LARS{} is to choose the unit vector $\bfu^{(k)}$ such that the active feature vectors also exhibit the same absolute correlation with the residual at the next step.
Note that we will see later in this section that $\sign(c^{(k+1)}_{i^*}) = \sign(c^{(k)}_{i^*})$ and thus $\bar{\bfX}^{(k+1)}_{i^*} = \bar{\bfX}^{(k)}_{i^*}$.
Consequently, the absolute correlations at the next step are $\bar{c}^{(k+1)}_{i^*} = \bar{\bfX}^{T(k)}_{i^*} \bfr^{(k+1)}_{\parallel} = \bar{c}^{(k)}_{i^*} - \gamma^{(k)} \bar{a}^{(k)}_{i^*}$, with $\bar{a}^{(k)}_{i^*}=\bar{\bfX}^{T(k)}_{i^*} \bfu^{(k)}$.
To achieve that the active feature vectors also exhibit the same absolute correlation with the residual at the next step, $\bar{a}^{(k)}_{i^*}$ must be equal for all $i^*\in\mathcal{A}^{(k)}$, i.e., the unit vector $\bfu^{(k)}$ must be chosen such that it is equiangular to all active feature vectors $\bar{\bfX}^{(k)}_{i^*}$.
The choice of the equiangular vector is not unique.
We are specifically interested in the equiangular vector that is in the span of $\bar{\bfX}^{(k)}_{i^*}$ and has a positive correlation with $\bar{\bfX}^{(k)}_{i^*}$.
To compute the equiangular vector $\bfu^{(k)}$, we define the matrix $\bar{\bfX}^{(k)}_{\mathcal{A}}$ which is composed of the columns $\bar{\bfX}^{(k)}_{i^*}$ for all $i^*\in\mathcal{A}^{(k)}$ and we define the vector of ones $\boldsymbol{1}_{\mathcal{A}}$.
The equiangular vector must fulfill the relationship $\bar{\bfX}^{T(k)}_{\mathcal{A}}\bfu^{(k)} = a \boldsymbol{1}_{\mathcal{A}}$ where $a$ is a scalar.
By multiplying $\boldsymbol{1}_{\mathcal{A}}$ with $\bar{\bfX}^{T(k)}_{\mathcal{A}}\bar{\bfX}^{(k)}_{\mathcal{A}}$ and its inverse, we obtain $\boldsymbol{1}_{\mathcal{A}}=\bar{\bfX}^{T(k)}_{\mathcal{A}}\bar{\bfX}^{(k)}_{\mathcal{A}}\left[\bar{\bfX}^{T(k)}_{\mathcal{A}}\bar{\bfX}^{(k)}_{\mathcal{A}}\right]^{-1}\boldsymbol{1}_{\mathcal{A}}$.
Thus, $\bar{\bfX}^{(k)}_{\mathcal{A}}\left[\bar{\bfX}^{T(k)}_{\mathcal{A}}\bar{\bfX}^{(k)}_{\mathcal{A}}\right]^{-1}\boldsymbol{1}_{\mathcal{A}}$ is an equiangular vector for $a=1$, which we normalize to obtain
\be
\label{eq:LARS_step_equiangular}
\bfu^{(k)} = A^{(k)} \bar{\bfX}^{(k)}_{\mathcal{A}}\left[\bar{\bfX}^{T(k)}_{\mathcal{A}}\bar{\bfX}^{(k)}_{\mathcal{A}}\right]^{-1}\boldsymbol{1}_{\mathcal{A}}
\quad
\text{with} \quad A^{(k)}
= \frac{1}{\| \bar{\bfX}^{(k)}_{\mathcal{A}}\left[\bar{\bfX}^{T(k)}_{\mathcal{A}}\bar{\bfX}^{(k)}_{\mathcal{A}}\right]^{-1}\boldsymbol{1}_{\mathcal{A}}\|}
= \frac{1}{\sqrt{\boldsymbol{1}^T_{\mathcal{A}}\left[\bar{\bfX}^{T(k)}_{\mathcal{A}}\bar{\bfX}^{(k)}_{\mathcal{A}}\right]^{-1}\boldsymbol{1}_{\mathcal{A}}}} > 0.
\ee
The inverse in the formula above is not computed explicitly. Instead, $\left[\bar{\bfX}^{T(k)}_{\mathcal{A}}\bar{\bfX}^{(k)}_{\mathcal{A}}\right]^{-1}\boldsymbol{1}_{\mathcal{A}}$ is computed by solving a linear system of equations.
We notice that $\bar{\bfX}^{T(k)}_{\mathcal{A}}\bfu^{(k)} = A^{(k)} \boldsymbol{1}_{\mathcal{A}} $, which means that the correlations $\bar{\bfX}^{T(k)}_{i^*} \bfu^{(k)} = A^{(k)}$ are positive, such that the angles between $\bar{\bfX}^{(k)}_{i^*}$ and $\bfu^{(k)}$ are smaller \mbox{than ${\pi}/{2}$}.

After computing the direction $\bfu^{(k)}$, we are left with identifying the step size $\gamma^{(k)}$.
The step size is always chosen positive, and it is chosen such that a new feature $\bfX_{j^*}$ enters the active set at the next step, i.e., $j^* \in \mathcal{A}^{\complement(k)}$, but $j^* \in \mathcal{A}^{(k+1)}$.
We are interested in the smallest possible positive step size, such that there exists an $\bfX_{j^*}$ with $|c^{(k+1)}_{j^*}| = \bar{c}^{(k+1)}_{\text{max}}$, i.e., either $c^{(k+1)}_{j^*} = \bar{c}^{(k+1)}_{\text{max}}$ or $-c^{(k+1)}_{j^*} = \bar{c}^{(k+1)}_{\text{max}}$.
The first condition leads to 
\be
\label{eq:LARS_step1}
    c^{(k+1)}_{j^*} = \bar{c}^{(k+1)}_{\text{max}}\quad
    \Rightarrow \quad c^{(k)}_{j^*} - \gamma^{(k)} a^{(k)}_{j^*} = \bar{c}^{(k)}_{\text{max}} - \gamma^{(k)} A^{(k)}
    \Rightarrow \quad \gamma^{(k)}
    = \frac{\bar{c}^{(k)}_{\text{max}} - c^{(k)}_{j^*}}{A^{(k)} - a^{(k)}_{j^*}},
\ee
while the second condition leads to
\be
\label{eq:LARS_step2}
\begin{aligned}
    -c^{(k+1)}_{j^*} = \bar{c}^{(k+1)}_{\text{max}}\quad
    \Rightarrow\quad & \gamma^{(k)}
    = \frac{\bar{c}^{(k)}_{\text{max}} + c^{(k)}_{j^*}}{A^{(k)} + a^{(k)}_{j^*}}. \\
\end{aligned}
\ee
Among all potential values for $\gamma^{(k)}$ in \cref{eq:LARS_step1,eq:LARS_step2}, we choose the smallest positive value
\be
\label{eq:LARS_stepsize}
\gamma^{(k)} = {\min_{j^* \in \mathcal{A}^{\complement(k)}}}^+ \left\{
\frac{\bar{c}^{(k)}_{\text{max}} - c^{(k)}_{j^*}}{A^{(k)} - a^{(k)}_{j^*}},
\frac{\bar{c}^{(k)}_{\text{max}} + c^{(k)}_{j^*}}{A^{(k)} + a^{(k)}_{j^*}}
\right\}.
\ee

After having identified $\bfu^{(k)}$ and $\gamma^{(k)}$, the prediction at the next step of \LARS{} is computed according to \cref{eq:LARS_step_prediction}.
Substituting $\bfu^{(k)}$ into \cref{eq:LARS_step_prediction} gives
\be
\bfmu^{(k+1)}=\bfmu^{(k)} + \gamma^{(k)} A^{(k)} \bar{\bfX}^{(k)}_{\mathcal{A}}\left[\bar{\bfX}^{T(k)}_{\mathcal{A}}\bar{\bfX}^{(k)}_{\mathcal{A}}\right]^{-1}\boldsymbol{1}_{\mathcal{A}}.
\ee
We split the prediction into the contributions from the inactive and active parameters $\bfmu^{(k)}=\bfX\bfw^{(k)}=\bfX_{\mathcal{A}^{\complement}}\bfw^{(k)}_{\mathcal{A}^{\complement}} + \bfX_{\mathcal{A}}\bfw^{(k)}_{\mathcal{A}}$.
Further, we use $\bar{\bfX}^{(k)}_{\mathcal{A}} = \bfX_{\mathcal{A}} \diag(\bfs^{(k)}_{\mathcal{A}})$ to arrive at
\be
\label{eq:LARS_step_parameters}
\begin{aligned}
\bfmu^{(k+1)}
&=\bfX_{\mathcal{A}^{\complement}}\bfw^{(k)}_{\mathcal{A}^{\complement}} + \bfX_{\mathcal{A}}\bfw^{(k)}_{\mathcal{A}} + \gamma^{(k)} A^{(k)} \bfX_{\mathcal{A}} \diag(\bfs^{(k)}_{\mathcal{A}})\left[\bar{\bfX}^{T(k)}_{\mathcal{A}}\bar{\bfX}^{(k)}_{\mathcal{A}}\right]^{-1}\boldsymbol{1}_{\mathcal{A}}\\
&=\bfX_{\mathcal{A}^{\complement}}\bfw^{(k)}_{\mathcal{A}^{\complement}} + \bfX_{\mathcal{A}}\Big[\bfw^{(k)}_{\mathcal{A}} + \underbrace{\gamma^{(k)} A^{(k)} \diag(\bfs^{(k)}_{\mathcal{A}})\left[\bar{\bfX}^{T(k)}_{\mathcal{A}}\bar{\bfX}^{(k)}_{\mathcal{A}}\right]^{-1}\boldsymbol{1}_{\mathcal{A}}}_{\Delta\bfw^{(k)}_{\mathcal{A}}}\Big],
\end{aligned}
\ee
where we identify the update of the active parameters $\Delta\bfw^{(k)}_{\mathcal{A}}$ in each step of \LARS{}.
See \cref{alg:LARS} for a summary of the \LARS{} algorithm.

\begin{algorithm}
\caption{\textit{Least Angle Regression} (LAR)}\label{alg:LARS}
\begin{algorithmic}
\State Given $\bfX$ and $\bfy$
\State Set $k=0$, $\bfw^{(0)}=\bf0$
\State $\bfy_{\parallel}=\bfX\left[\bfX^T\bfX\right]^{-1}\bfX^T\bfy$ 
\While{$\|\bfw^{(k)}\|_0 < m-1$}
\State $\bfc^{(k)} = \bfX^T [\bfy_{\parallel} - \bfX\bfw^{(k)}]$
\State $\mathcal{A}^{(k)} = \left\{ i^*\in \{1,\dots,m\} ~ \big| ~ |c^{(k)}_{i^*}| = \bar{c}^{(k)}_{\text{max}} = \max_i |c^{(k)}_i| \right\}$, $\mathcal{A}^{\complement(k)} = \left\{ j^*\in \{1,\dots,m\} ~ \big| ~ j^* \notin \mathcal{A}^{(k)} \right\}$
\State $\bfs^{(k)}=\sign(\bfc^{(k)})$
\State $\bar{\bfX}^{(k)}_{\mathcal{A}} = \bfX_{\mathcal{A}} \diag(\bfs^{(k)}_{\mathcal{A}})$
\State $A^{(k)} = 1/\sqrt{\boldsymbol{1}^T_{\mathcal{A}}\left[\bar{\bfX}^{T(k)}_{\mathcal{A}}\bar{\bfX}^{(k)}_{\mathcal{A}}\right]^{-1}\boldsymbol{1}_{\mathcal{A}}}$
\State $\bfu^{(k)} = A^{(k)} \bar{\bfX}^{(k)}_{\mathcal{A}}\left[\bar{\bfX}^{T(k)}_{\mathcal{A}}\bar{\bfX}^{(k)}_{\mathcal{A}}\right]^{-1}\boldsymbol{1}_{\mathcal{A}}$
\State $\gamma^{(k)} = \min_{j^* \in \mathcal{A}^{\complement(k)}}^+ \left\{
\frac{\bar{c}^{(k)}_{\text{max}} - c^{(k)}_{j^*}}{A^{(k)} - a^{(k)}_{j^*}},
\frac{\bar{c}^{(k)}_{\text{max}} + c^{(k)}_{j^*}}{A^{(k)} + a^{(k)}_{j^*}}
\right\}$
\State $\bfw^{(k+1)}_{\mathcal{A}^{\complement}} \gets \bfw^{(k)}_{\mathcal{A}^{\complement}}$
\State $\bfw^{(k+1)}_{\mathcal{A}} \gets \bfw^{(k)}_{\mathcal{A}} + \gamma^{(k)} A^{(k)} \diag(\bfs^{(k)}_{\mathcal{A}})\left[\bar{\bfX}^{T(k)}_{\mathcal{A}}\bar{\bfX}^{(k)}_{\mathcal{A}}\right]^{-1}\boldsymbol{1}_{\mathcal{A}}$
\State $k \gets k+1$
\EndWhile
\State $\bfw^{(k+1)} \gets \left[\bfX^T\bfX\right]^{-1}\bfX^T\bfy$
\end{algorithmic}
\end{algorithm}

Choosing the step size as shown in \cref{eq:LARS_stepsize} has an interesting effect on the evolution of correlations corresponding to the active set over each step of \LARS{}.
Specifically, we recall that $c^{(k+1)}_{i^*} = c^{(k)}_{i^*} - \gamma^{(k)} \bfX_{i^*}^T \bfu^{(k)} = \sign(c^{(k)}_{i^*}) [\bar{c}^{(k)}_{\text{max}} - \gamma^{(k)} A^{(k)}] $.
From \cref{eq:LARS_stepsize}, we deduce that $\gamma^{(k)} A^{(k)} < \bar{c}^{(k)}_{\text{max}}$, see \cref{sec:bounded_step_size}, which means that the signs of the correlations corresponding to the active set do not change over one step, $\sign(c^{(k+1)}_{i^*}) = \sign(c^{(k)}_{i^*})$.
And at each step, the maximum absolute correlation decreases according to $|c^{(k+1)}_{i^*}| = \bar{c}^{(k)}_{\text{max}} -  \gamma^{(k)} A^{(k)}$.

\subsubsection{\LARSLASSO{}}
\label{sec:LARSLASSO}

The previously described \LARS{} algorithm provides an efficient tool for model discovery.
The solutions $\bfw^{(k)}$ computed by \LARS{} are similar to those obtained by solving \refp{prob:1} for different values of $\alpha$.
However, under certain conditions, \LARS{} can yield solutions that cannot be identified as solutions of \refp{prob:1}.
Specifically, it can be shown \citep{efron_least_2004} that any nonzero parameter $w^*_i \neq 0$ of a solution to \refp{prob:1} must fulfill $\sign(w^*_i) = \sign(c_i)$ with $c_i=\bfX_i^T [\bfy_{\parallel} - \bfX\bfw^*]$,
and \LARS{} can yield solutions that violate this condition.
Thus, \cite{efron_least_2004} proposed a modification of \LARS{}, which we call \LARSLASSO{}, that is specifically designed to find solutions to \refp{prob:1}.

The algorithm starts off by initializing all parameters to zero $\bfw^{(0)}=\bf0$.
This is a valid solution to \refp{prob:1} for a sufficiently large value of $\alpha$, as discussed in \cref{sec:boundedness_regularization_parameter}.
At each step, we assume that $\sign(w^{(k)}_i) = \sign(c^{(k)}_i)$ for $w^{(k)}_i \neq 0$, and we modify the \LARS{} steps, such that, after each step, the condition $\sign(w^{(k+1)}_i) = \sign( c^{(k+1)}_i)$ for $w^{(k+1)}_i \neq 0$ is fulfilled.
As discussed after \cref{eq:LARS_step_equiangular}, the signs of the correlations corresponding to the active set do not change over one step.
We distinguish two scenarios:
First, we consider the case where $w^{(k)}_i = 0$ and $w^{(k+1)}_i \neq 0$.
For this case, it is $\sign(w^{(k+1)}_i) = \sign( c^{(k)}_i)$, see \cref{sec:enter_the_active_set}, and therefore $\sign(w^{(k+1)}_i) = \sign( c^{(k+1)}_i)$, which means that the condition is satisfied.
Second, the condition may be violated if $w^{(k)}_i \neq 0$ and $\sign(w^{(k+1)}_i) \neq \sign(w^{(k)}_i)$.
We recall the update of the active parameters in each step, see \cref{eq:LARS_step_parameters},
\be
\bfw^{(k+1)}_{\mathcal{A}} = \bfw^{(k)}_{\mathcal{A}} + \Delta\bfw^{(k)}_{\mathcal{A}} = \bfw^{(k)}_{\mathcal{A}} + \gamma^{(k)} A^{(k)} \diag(\bfs^{(k)}_{\mathcal{A}})\left[\bar{\bfX}^{T(k)}_{\mathcal{A}}\bar{\bfX}^{(k)}_{\mathcal{A}}\right]^{-1}\boldsymbol{1}_{\mathcal{A}},
\ee
which we write in index notation as
\be
w^{(k+1)}_{i^*} = w^{(k)}_{i^*} + \Delta w^{(k)}_{i^*} = w^{(k)}_{i^*} + \gamma^{(k)} d^{(k)}_{i^*},
\ee
where $i^*\in\mathcal{A}^{(k)}$ and $d^{(k)}_{i^*}$ are the associated entries of the vector $A^{(k)} \diag(\bfs^{(k)}_{\mathcal{A}})\left[\bar{\bfX}^{T(k)}_{\mathcal{A}}\bar{\bfX}^{(k)}_{\mathcal{A}}\right]^{-1}\boldsymbol{1}_{\mathcal{A}}$.
We recall that $\gamma^{(k)} > 0$ and observe that the sign of the parameters changes if and only if $\sign(w^{(k)}_{i^*}) \neq \sign(d^{(k)}_{i^*})$ and $\gamma^{(k)} > - w^{(k)}_{i^*}/d^{(k)}_{i^*}$ for any $i^*\in\mathcal{A}^{(k)}$.
We define $\tilde\gamma = \min_{i^*\in\mathcal{A}^{(k)}}^+ \{ - w^{(k)}_{i^*}/d^{(k)}_{i^*} \}$ as the smallest possible positive value of $\gamma^{(k)}$ for which one of the parameters switches its sign.
For the special case that $\sign(w^{(k)}_{i^*}) = \sign(d^{(k)}_{i^*})$ for all $i^*\in\mathcal{A}$, we set $\tilde\gamma = \infty$.
We finally modify the update rule such that, if the stepsize $\gamma^{(k)}$ computed by \eqref{eq:LARS_stepsize} is greater than $\tilde\gamma$, we set instead $\gamma^{(k)} = \tilde\gamma$.
In this way, the parameter $w^{(k)}_{\tilde i}$ with $\tilde i\in\mathcal{A}^{(k)}$ that would first switch its sign upon increasing $\gamma^{(k)}$ equates to zero after the step $w^{(k+1)}_{\tilde i}=0$.
Note that we follow \cite{efron_least_2004} and assume that for $\gamma^{(k)} = \tilde\gamma$, only one parameter equates to zero.
The index $\tilde i$ that corresponds to this parameter is excluded from the active set at the next step. 

Each solution $\bfw^{(k)}$ obtained by \LARSLASSO{} is a solution to \refp{prob:1} \citep{efron_least_2004}.
Given the solution $\bfw^{(k)}$, the corresponding regularization parameter $\alpha^{(k)}$ can be computed as detailed in \ref{sec:regularization_parameter_LARSLASSO}.
Specifically, we obtain
\be
\alpha^{(k)} = \frac{\bar{c}^{(k)}_{\text{max}}}{n}.
\ee
The modified algorithm \LARSLASSO{} is summarized in \cref{alg:LARSLASSO} in \cref{sec:implementation}.

\LARSLASSO{} can be interpreted as a \textit{bottom-up} approach, as it starts from the zero solution and successively adds nonzero components to the parameter vector.
As discussed in \cref{sec:results}, when discovering models to describe the mechanical behavior of materials, we are typically interested in models with a small number of material parameters, often as few as two or three for isotropic materials.
This makes \textit{bottom-up} approaches more favorable in practice, as they are expected to identify the practically relevant models more efficiently than \textit{top-down} methods.

\LARSLASSO{} efficiently solves \refp{prob:2}, however, we note that \LARSLASSO{} may also be used to efficiently solve \refp{prob:1}.
If the critical values of the regularization parameters and the corresponding parameters $\bfw^*(\alpha_c)$ are known, see \refp{prob:2}, solutions for regularization parameters between critical values can be obtained through linear interpolation between the parameters $\bfw^*(\alpha_c)$.
\LARSLASSO{} is computationally more efficient than CD, especially if we are only interested in the large regularization parameter regime, i.e., the first steps of \LARSLASSO{}.


\subsection{\textit{Iterative Soft-Thresholding Algorithm} (ISTA)}

We finally move on to models that depend nonlinearly on the parameters and discuss strategies for numerically solving \refp{prob:3}.
Specifically, we put our attention on the \textit{Iterative Soft-Thresholding Algorithm} (ISTA) which is a first-order method belonging to the family of \textit{proximal gradient methods} \citep{parikh_proximal_2013,beck_first-order_2017}.
ISTA is shown in \cref{alg:ISTA} and briefly described in the following.
For a detailed description and a mathematical treatment of the method, we refer to \cite{beck_first-order_2017}.

\begin{algorithm}
\caption{\textit{Iterative Soft-Thresholding Algorithm} (ISTA)}\label{alg:ISTA}
\begin{algorithmic}
\State Given $f$
\State Set initial guess $\bfw^{(0)}$
\State Choose the maximum number of steps NSTEP and convergence tolerance TOL
\For{$k=0,\dots,\text{NSTEP}-1$}
\State $\bfw^{(k+1)} = \text{prox}_{\gamma g}(\bfw^{(k)} - \gamma\nabla f(\bfw^{(k)}))$
\If{$\| \bfw^{(k+1)} - \bfw^{(k)} \|_2 < \text{TOL}$ \textbf{or} $| f(\bfw^{(k+1)}) + g(\bfw^{(k+1)}) - f(\bfw^{(k)}) - g(\bfw^{(k)}) | < \text{TOL}$}
\State \textbf{break}
\EndIf
\EndFor
\end{algorithmic}
\end{algorithm}

\CHANGE{
\refp{prob:3} can be mathematically interpreted as a so-called composite problem \citep{beck_first-order_2017}, a multi-objective optimization problem, or, more simply, a regularized optimization problem
\be
\bfw^* = \argmin_{\bfw} \left[ f(\bfw) + g(\bfw) \right],
\ee
where for our case $g(\bfw) = \alpha \|\bfw\|_1$.
}
\CHANGE{
\textit{Proximal gradient methods} constitute a family of algorithms designed to solve composite problems.
}
They are first order methods relying on gradient computations of $f(\bfw)$ and thus share similarities with classical gradient descent algorithms.
They start off with an initial guess $\bfw^{(0)}$ and evaluate the gradient $\nabla f(\bfw^{(k)}) = {\partial f}/{\partial \bfw} (\bfw^{(k)})$ at each iteration $k$.
As in classical gradient descent algorithms, the current solution is updated by making a step of a given step size $\gamma > 0$ into the direction of the negative gradient, i.e., $\bfw^{(k)} - \gamma\nabla f(\bfw^{(k)})$.
Afterwards, and in contrast to classical gradient descent algorithms, a so-called proximal mapping $\text{prox}_{\gamma g}(\cdot)$ is applied, such that one step of the \textit{proximal gradient methods} can be summarized as
\be
\label{eq:step_ISTA}
\bfw^{(k+1)} = \text{prox}_{\gamma g}(\bfw^{(k)} - \gamma\nabla f(\bfw^{(k)})).
\ee
The proximal mapping or proximal operator of a function $h(\bfw)$ is defined through
\be
\text{prox}_{h}(\bfw) = \argmin_{\bfu} \left[ h(\bfu) + \frac{1}{2} \| \bfu - \bfw \|_2^2 \right].
\ee
Thus, the proximal mapping $\text{prox}_{\gamma g}(\bfw^{(k)})$ in \cref{eq:step_ISTA} with $g(\bfw) = \alpha \|\bfw\|_1$ is
\be
\text{prox}_{\gamma g}(\bfw) = \argmin_{\bfu} \left[ \gamma \alpha \|\bfu\|_1 + \frac{1}{2} \| \bfu - \bfw \|_2^2 \right].
\ee
We notice the similarity between the minimization problem above and \refp{prob:1} with the feature matrix being the identity $\bfX=\bfI$. 
This problem has a closed-form solution, which is written in index notation as \citep{beck_first-order_2017}
\be
\{\text{prox}_{\gamma g}(\bfw)\}_i = \text{soft}_{\gamma \alpha}(w_i) = \sign(w_i) \max \{ |w_i| - \gamma \alpha , 0 \} =
\begin{cases}
    w_i - \gamma \alpha & \text{if } w_i > \gamma \alpha \\
    0 & \text{if } - \gamma \alpha \leq w_i \leq \gamma \alpha \\
    w_i + \gamma \alpha & \text{if } w_i < -\gamma \alpha \\
\end{cases},
\ee
which we identify as the soft-thresholding function similar to \cref{eq:CD_step_closed_form,eq:CD_step_closed_form2}.

The convergence proof of ISTA for solving \refp{prob:3} with $\nabla f(\bfw)$ being Lipschitz continuous can be found in \cite{beck_first-order_2017}.
We note that the choice of the step size directly influences the convergence behavior of ISTA.
If the step size is chosen too large, ISTA does not converge, and if the step size is chosen too small, the number of iterations to achieve convergence increases.
If the Lipschitz constant of $\nabla f(\bfw)$ is known, the step size can be chosen dependent on the Lipschitz constant \citep{beck_first-order_2017}.
In practice, however, the Lipschitz constant is usually not known such that a suitable step size must be determined through trial and error. 
Obviously, ISTA can also be applied to \refp{prob:1}.
However, CD typically outperforms ISTA in terms of computational efficiency for \refp{prob:1}.
Further, CD is preferred over ISTA because it does not require choosing a step size.

We note that, dependent on the initial guess, ISTA can be interpreted as either a \textit{bottom-up} approach or a \textit{top-down} approach.
If the zero vector is used as the initial guess, ISTA successively adds more nonzero parameters to the solution.
In contrast, if a dense vector is used as the initial guess, ISTA progressively sets more and more parameters to zero.

\subsection{Pathwise ISTA}
ISTA solves \refp{prob:3} for a given value of $\alpha$.
However, to select a suitable value of $\alpha$, it is often beneficial to compute the regularization path of \refp{prob:3}.
Computing the regularization path for models that depend nonlinearly on the parameters is not straightforward.
Yet, in the literature, different methods for computing approximations of the regularization path have been proposed \citep{friedman_pathwise_2007,friedman_regularization_2010,yang_fast_2024-1,yang_fast_2024}.
These methods start off with a value of $\alpha$ that yields the zero solution and then successively decrease $\alpha$ in predefined step sizes to approximately compute the regularization path, see \refp{prob:4}.
Importantly, solutions from previous computations serve as initial guesses for subsequent computations, which decreases the computational costs significantly.
While the aforementioned works focus on second order optimization methods, in the following, we apply the same philosophy to the first order method ISTA and develop a pathwise ISTA.

\begin{algorithm}
\caption{\textit{Pathwise Iterative Soft-Thresholding Algorithm} (Pathwise ISTA)}\label{alg:pathwise_ISTA}
\begin{algorithmic}
\State Given $f$ and $n_\alpha$
\State Choose the maximum number of steps NSTEP and convergence tolerance TOL
\State $\alpha^{(0)} = \max_i \left|\frac{\partial f}{\partial w_i}(\boldsymbol{0})\right|$
\State $\alpha^{(l)} = (1 - \frac{l}{n_\alpha}) \alpha^{(0)}$
\State $\bfw^{(0)} = \boldsymbol{0}$
\For{$l=1,\dots,n_\alpha-1$}
\State $\alpha = \alpha^{(l)}$
\State $\bfw^{(l)(0)} = \bfw^{(l-1)}$
\For{$k=0,\dots,\text{NSTEP}-1$}
\State $\bfw^{(l)(k+1)} = \text{prox}_{\gamma g}(\bfw^{(l)(k)} - \gamma\nabla f(\bfw^{(l)(k)}))$
\If{$\| \bfw^{(l)(k+1)} - \bfw^{(l)(k)} \|_2 < \text{TOL}$ \textbf{or} $| f(\bfw^{(l)(k+1)}) + g(\bfw^{(l)(k+1)}) - f(\bfw^{(l)(k)}) - g(\bfw^{(l)(k)}) | < \text{TOL}$}
\State $\bfw^{(l)} = \bfw^{(l)(k+1)}$
\State \textbf{break}
\EndIf
\EndFor
\EndFor
\end{algorithmic}
\end{algorithm}

\cref{alg:pathwise_ISTA} details the functionality of the pathwise ISTA.
The pathwise ISTA is a \textit{bottom-up} approach and starts off with $\bfw^{(0)} = \boldsymbol{0}$.
As shown in \cref{sec:boundedness_regularization_parameter}, $\bfw^{(0)} = \boldsymbol{0}$ is a stationary point of the underlying minimization problem for $\alpha^{(0)} = \max_i \left|\frac{\partial f}{\partial w_i}(\boldsymbol{0})\right|$.
In each step $l=1,\dots,n_\alpha-1$ of the pathwise ISTA, we successively decrease the regularization parameter according to $\alpha^{(l)} = (1 - \frac{l}{n_\alpha}) \alpha^{(0)}$ and solve the minimization problem using ISTA.
At each \mbox{step $l$}, ISTA computes a sequence of solutions $\bfw^{(l)(k)}$ using the initial guess $\bfw^{(l)(k)} = \bfw^{(l-1)}$ until a convergence criterion is met.
The converged solution $\bfw^{(l)}$ constitutes the solution corresponding to the regularization parameter $\alpha^{(l)}$ and serves as the initial guess for the subsequent step.

\section{Automated material model discovery}
\label{sec:material_model_discovery}
The mathematical problems presented in \cref{sec:mathematical_problems} constitute the backbone of the broad field of library-based material model discovery \citep{flaschel_unsupervised_2021,wang_inference_2021,wang_establish_2022,linka_new_2023,meyer_thermodynamically_2023,fuhg_extreme_2024,moon_physics-informed_2025}.
In the following, we draw a link to the problems in \cref{sec:mathematical_problems} and automated material model discovery by introducing several example problems.
In particular, we detail how the model-data-mismatch $f(\bfw)$ can be formulated in the context of material modeling.
We focus our attention on incompressible hyperelastic material models that either depend linearly or nonlinearly on the material parameters, see \cite{marckmann_comparison_2006,chagnon_hyperelastic_2015,dal_performance_2021} for recent reviews, and consider labeled data from experiments with homogeneous deformation fields such as uniaxial tension or simple shear.

\subsection{Material model library}
\label{sec:material_model_library}
In this work, we focus on library-based approaches for material model discovery, which constitute at the time the most prominent methods for material model discovery, see for example the works by \cite{flaschel_unsupervised_2021,wang_inference_2021,wang_establish_2022,linka_new_2023,meyer_thermodynamically_2023,fuhg_extreme_2024,moon_physics-informed_2025}.
That is, we formulate a general parametric ansatz for the material model and use the previously described methods to identify which of the parameters in the ansatz are necessary to describe the given data and which of the parameters may be set to zero.
By identifying the most important parameters and setting others to zero, we arrive at a concise and thus interpretable mathematical expression of the material model.

A material model library for incompressible hyperelastic materials can be formulated by introducing a general parametric ansatz for the material's strain energy density function.
Under the assumption of incompressibility and isotropy, the strain energy density $W$ of a hyperelastic material is postulated as
\be
W = \tilde{W}(I_1,I_2,\lambda_1,\lambda_2,\lambda_3;\bfw) - p \cdot [J-1],
\ee
where $I_1=\text{tr} (\bfC)$, $I_2=\frac{1}{2}(\text{tr}^2 (\bfC) - \text{tr} (\bfC^2))$ are invariants of the right Cauchy-Green stretch tensor $\bfC=\bfF^T\bfF$, $\bfF$ is the deformation gradient, $\lambda_1$, $\lambda_2$, $\lambda_3$ are the principal stretches defined as the eigenvalues of the right stretch tensor $\bfU$ obtained through the polar decomposition $\bfF=\bfR\bfU$, $p$ is a scalar Lagrange multiplier that can be physically interpreted as the pressure, and $J=\det\bfF\overset{!}{=}1$ is the determinant of the deformation gradient \citep{holzapfel_nonlinear_2000}.
Our objective is to discover the function $\tilde{W}(I_1,I_2,\lambda_1,\lambda_2,\lambda_3;\bfw)$ and its unknown parameters $\bfw$.
Note that we explicitly indicate the dependence of the energy on both the invariants and the principal stretches, even though the invariants can be expressed as functions of the principal stretches, because we seek to design algorithms to automatically discover whether invariant-based or principal-stress-based models are superior to describe a given dataset.

The general ansatz $\tilde{W}(I_1,I_2,\lambda_1,\lambda_2,\lambda_3;\bfw)$ comprises many well-known phenomenological material models, such as, for example, Mooney-Rivlin-type models or Ogden-type models.
Existing material models can be broadly classified into models that depend linearly or nonlinearly on the material parameters $\bfw$.
The classification of these models determines the structure of the optimization problem for determining the unknown parameters, see \refp{prob:1}, \refp{prob:2} for the linear case and \refp{prob:3} for the nonlinear case.
Therefore, in the following, we treat these two cases separately and introduce two different material model libraries.

\subsubsection{Linear material model library}
\label{sec:material_model_library_linear}
Assuming that the model depends linearly on the parameters, the strain energy density may be written as
\be
\tilde{W}(I_1,I_2,\lambda_1,\lambda_2,\lambda_3;\bfw) = \bfQ(I_1,I_2,\lambda_1,\lambda_2,\lambda_3)^T \bfw,
\ee
where $\bfQ \in \Rset^m$ is a vector containing feature functions that depend on the invariants and principal stretches.
We emphasize that, although $\tilde{W}$ depends linearly on $\bfw$, the feature functions are nonlinear in general, which means that the model is still able to describe the nonlinear material responses of hyperelastic materials.

In this work, we consider the invariant-based generalized Mooney-Rivlin model \citep{rivlin_torsion_1947,rivlin_large_1950,rivlin_large_1951} as the material model library
\be
\tilde{W}(I_1,I_2;\bfw) = \sum_{i=1}^{n_{\text{Mooney}}} \sum_{j=0}^i C_{ij} (I_1 - 3)^{i-j} (I_2 - 3)^j,
\ee
where $n_{\text{Mooney}}$ defines the maximum polynomial order, and the material parameters $C_{ij}$ can be collected into a vector $\bfw$ such that $\tilde{W}(I_1,I_2;\bfw) = \bfQ(I_1,I_2)^T \bfw$ with $\bfQ$ being a vector containing the nonlinear feature functions \mbox{$(I_1 - 3)^{i-j} (I_2 - 3)^j$}.
The identification of the parameters in this model, without considering the $L_1$-regularization, has been comprehensively studied by \cite{hartmann_parameter_2001}.

\subsubsection{Nonlinear material model library}
\label{sec:material_model_library_nonlinear}

Some of the well-known material models used in practice depend nonlinearly on the material parameters, such as for example Ogden-type material models \citep{ogden_large_1972}.
Thus, we consider the general material model library
\be
\tilde{W}(I_1,I_2,\lambda_1,\lambda_2,\lambda_3;\bfw) =
\sum_{i=1}^{n_{\text{Mooney}}} \sum_{j=0}^i C_{ij} [I_1 - 3]^{i-j} [I_2 - 3]^j
+ D
\left[\lambda_1^{\delta} + \lambda_2^{\delta} + \lambda_3^{\delta} - 3\right],
\ee
which includes the previously described modeling features and an additional Ogden-type feature, which depends nonlinearly on the material parameters, see also \cite{flaschel_automated_2023-1}. The material parameter vector now comprises the parameters $C_{ij}$, $D$ and $\delta$.

\subsection{Stress-strain relationship}
The relationship between deformation and stress is obtained by differentiating the strain energy density.
Specifically, the Piola stress $\bfP$ is computed by differentiating $W$ with respect to the deformation gradient
\be
\label{eq:stress}
\bfP
= \frac{\partial W}{\partial \bfF}
= \frac{\partial \tilde{W}}{\partial \bfF} - p\bfF^{-T},
\ee
where we used $\frac{\partial J}{\partial \bfF} = J\bfF^{-T}$.
Applying the chain rule, we obtain
\be\label{eq:stress_chain_rule}
\frac{\partial \tilde{W}}{\partial F_{ij}}
= \frac{\partial \tilde{W}}{\partial I_a}\frac{\partial I_a}{\partial F_{ij}}
+ \frac{\partial \tilde{W}}{\partial \lambda_b}\frac{\partial \lambda_b}{\partial F_{ij}}.
\ee
The derivatives of the strain energy density with respect to the strain invariants and the principal stretches can be computed analytically, 
or by means of automatic differentiation.
Assuming $\lambda_1\neq\lambda_2\neq\lambda_3\neq\lambda_1$, the derivatives of the strain invariants and the principal stretches with respect to the deformation gradient are
\be
\frac{\partial I_1}{\partial \bfF} = 2\bfF, \quad
\frac{\partial I_2}{\partial \bfF} = 2I_1\bfF - 2\bfF\bfC, \quad
\frac{\partial \lambda_1}{\partial \bfF} = \bfn_1\otimes\bfN_1, \quad
\frac{\partial \lambda_2}{\partial \bfF} = \bfn_2\otimes\bfN_2, \quad
\frac{\partial \lambda_3}{\partial \bfF} = \bfn_3\otimes\bfN_3,
\ee
where $\bfN_i$ denote the eigenvectors of $\bfC=\bfF^T\bfF$ and $\bfn_i$ the eigenvectors of $\bfb=\bfF\bfF^T$ \citep{holzapfel_nonlinear_2000}.

We focus on experiments with simple deformation fields for measuring stress-strain data pairs.
Specifically, we put our attention on uniaxial compression/tension and simple shear.
As we discuss in the following, under these load cases, the deformation gradient follows a specific structure, which simplifies the stress-strain relationship.

\subsubsection{Uniaxial compression and tension}
Experimental measurements of specimens under uniaxial compression and tension deliver labeled data pairs in the form $(F_{11},P_{11})$, where $F_{11}$ and $P_{11}$ are the longitudinal normal components of the deformation gradient and the Piola stress, respectively.
As a result of the incompressibility assumption, $\det\bfF\overset{!}{=}1$, and due to the symmetry condition, $F_{22}=F_{33}$, the deformation gradient under uniaxial compression/tension reads
\be
\begin{aligned}
\bfF =
	\begin{bmatrix}
		F_{11} & 0 & 0\\
		0 & \frac{1}{\sqrt{F_{11}}} & 0\\
		0 & 0 & \frac{1}{\sqrt{F_{11}}}\\
	\end{bmatrix}.
\end{aligned}
\ee
To obtain a relationship between the longitudinal normal components of the deformation gradient and the Piola stress $P_{11}(F_{11};\bfw)$, the unknown hydrostatic pressure $p$ in \cref{eq:stress} needs to be computed.
Using the zero-normal-stress condition,
$P_{22}=P_{33}\overset{!}{=}0$
along with \cref{eq:stress}, we find the hydrostatic pressure
\be
P_{33} = \frac{\partial \tilde{W}}{\partial F_{33}}- pF^{-1}_{33} \overset{!}{=} 0, \quad \Rightarrow \quad p = \frac{\partial \tilde{W}}{\partial F_{33}} F_{33}.
\ee
We hence obtain the desired relationship by substituting the pressure in \cref{eq:stress}
\be
\label{eq:constitutive_map_uniaxial_tension}
P_{11}(F_{11};\bfw)
= \frac{\partial \tilde{W}}{\partial F_{11}} - pF^{-1}_{11}
= \frac{\partial \tilde{W}}{\partial F_{11}} - \frac{F_{33}}{F_{11}}\frac{\partial \tilde{W}}{\partial F_{33}}.
\ee

\subsubsection{Simple shear}
Experimental measurements of specimens under simple shear deliver labeled data pairs in the form $(F_{12},P_{12})$, where $F_{12}$ and $P_{12}$ are the shear components of the deformation gradient and the Piola stress, respectively.
The deformation gradient under simple shear simplifies to
\be
\begin{aligned}
\bfF =
	\begin{bmatrix}
		1 & F_{12} & 0\\
		0 & 1 & 0\\
		0 & 0 & 1\\
	\end{bmatrix},
\quad
\bfF^{-T} =
	\begin{bmatrix}
		1 & 0 & 0\\
		-F_{12} & 1 & 0\\
		0 & 0 & 1\\
	\end{bmatrix},
\end{aligned}
\ee
which satisfies the incompressibility constraint. Because the shear component of the transposed inverse of the deformation gradient vanishes $\{\bfF_{\text{SS}}^{-T}\}_{12} = 0$, the relationship $P_{12}(F_{12};\bfw)$ does not depend on the pressure $p$, and we obtain
\be
\label{eq:constitutive_map_simple_shear}
P_{12}(F_{12};\bfw) = \frac{\partial \tilde{W}}{\partial F_{12}}.
\ee

\subsection{Model-data-mismatch}

To infer information about the material parameters $\bfw$, we base in this work on uniaxial tension/compression data in the form of data pairs $(F_{11}^{(i)},P_{11}^{(i)})$ with $i=1,\dots,n_{\text{UTC}}$, while the simple shear data takes the form $(F_{12}^{(j)},P_{12}^{(j)})$ with $j=1,\dots,n_{\text{SS}}$.
We define $P_{11}^{\text{max}} = \max_{i} |P_{11}^{(i)}|$ and $P_{12}^{\text{max}} = \max_{j} |P_{12}^{(j)}|$ and choose the model-data-mismatch for the nonlinear material model library in \cref{sec:material_model_library_nonlinear} as
\be
f(\bfw) =
\frac{1}{2[n_{\text{UTC}} + n_{\text{SS}}]} \left[ \sum_{i=1}^{n_{\text{UTC}}} \left[ \frac{P_{11}(F_{11}^{(i)};\bfw) - P_{11}^{(i)}}{P_{11}^{\text{max}}} \right]^2
+ \sum_{i=1}^{n_{\text{SS}}} \left[ \frac{P_{12}(F_{12}^{(i)};\bfw) - P_{12}^{(i)}}{P_{12}^{\text{max}}} \right]^2 \right] ,
\ee
which provides a metric for quantifying the mismatch between the model predictions and the data.
Due to the division by $P_{ij}^{\text{max}}$, the model-data-mismatch is non-dimensionalized.
In this way, the contributions from the uniaxial tension/compression data and the simple shear data exert an equal influence on the model–data mismatch.

The model-data-mismatch for the linear material model library in \cref{sec:material_model_library_linear} is defined similarly.
However, we additionally consider a normalization of the feature vectors.
To this end, we assemble the feature matrix
\be
\tilde\bfX = 
\begin{bmatrix}
    \tilde\bfX^{\text{UTC}} \\
    \tilde\bfX^{\text{SS}} \\
\end{bmatrix}, \quad \text{with} \quad
\tilde{X}^{\text{UTC}}_{ij} = \frac{1}{P_{11}^{\text{max}}} \left[ \frac{\partial Q_j}{\partial F_{11}}(F^{(i)}_{11}) - \frac{F^{(i)}_{33}}{F^{(i)}_{11}}\frac{\partial Q_j}{\partial F_{33}}(F^{(i)}_{11}) \right] , \quad
\tilde{X}^{\text{SS}}_{ij} = \frac{1}{P_{12}^{\text{max}}} \frac{\partial Q_j}{\partial F_{12}}(F^{(i)}_{12}),
\ee
where we considered \cref{eq:constitutive_map_uniaxial_tension,eq:constitutive_map_simple_shear},
and the measurement vector
\be
\bfy = 
\begin{bmatrix}
    \bfy^{\text{UTC}} \\
    \bfy^{\text{SS}} \\
\end{bmatrix}, \quad \text{with} \quad
{y}^{\text{UTC}}_i = \frac{P_{11}^{(i)}}{P_{11}^{\text{max}}}, \quad
{y}^{\text{SS}}_i = \frac{P_{12}^{(i)}}{P_{12}^{\text{max}}}.
\ee
Then, we assemble a normalized feature matrix $\bfX$, whose columns are computed as $\bfX_i = \tilde\bfX_i / \|\tilde\bfX_i\|_2$, see \cref{sec:mathematical_problems_1}.
The model-data-mismatch for the linear material model library is finally defined as 
\be
f(\bfw) =
\frac{1}{2[n_{\text{UTC}} + n_{\text{SS}}]} || \bfy - \bfX \bfw ||^2,
\ee
where we note that, after using this model-data-mismatch for CD or \LARSLASSO{}, the material parameters must be rescaled according to $\tilde\bfw_i = \bfw_i / \|\tilde\bfX_i\|_2$, in which $\tilde\bfw_i$ are the actual material model parameters.
\section{Benchmarking}
\label{sec:results}

In the following, we apply the discussed algorithms to different benchmark problems.
We will distinguish between the material models that depend linearly and nonlinearly on the material parameters.

\subsection{Linear material model library}
\label{sec:results_linear}

First, we consider material models that depend linearly on the material parameters.
We consider four benchmark material models; the Neo-Hookean model, the Mooney-Rivlin model, the Yeoh model, and the Biderman model.
The models' strain energy density functions and the values of the material parameters are shown in \cref{tab:benchmarks_linear}.
We generate synthetic data for the four material models by choosing $n_{\text{UTC}}$ equidistant values between $0.75$ and $1.5$ for $F_{11}^{(i)}$ in the uniaxial tension/compression case, and $n_{\text{SS}}$ equidistant values between $0$ and $0.5$ for $F_{12}^{(j)}$ in the simple shear case.
We consider both the noise-free data as well as data perturbed by independent Gaussian noise, i.e., $P_{ij \ \text{noisy}}^{(i)} = P_{ij}^{(i)} + \varepsilon$ with $\varepsilon \sim \mathcal{N}(0,\sigma)$.
Specifically, we choose a standard deviation of $\sigma=5$, resulting in an exceptionally high noise in the data.


\begin{table}[h!]
\caption{Strain energy density functions of the true and discovered material models and model-data-mismatch.}
\label{tab:benchmarks_linear}
\centering
\begin{tabular}{|ll|l|r|}
\hline
\multicolumn{2}{|l|}{Benchmarks\vphantom{$\frac{\int}{a}$}} & \multicolumn{1}{c|}{Strain energy density $\tilde{W}$} & \multicolumn{1}{c|}{$f(\bfw)$} \\ \hline
\multicolumn{1}{|l|}{Neo-Hookean\vphantom{$\frac{\int}{a}$}} & Truth & $40.00\left[I_1 - 3\right]$ & - \\ \cline{2-4} 
\multicolumn{1}{|l|}{\vphantom{$\frac{\int}{a}$}}            & $\sigma=0$ & $40.00\left[I_1 - 3\right]$ & $7.34 \cdot 10^{-33}$ \\ \cline{2-4} 
\multicolumn{1}{|l|}{\vphantom{$\frac{\int}{a}$}}            & $\sigma=5$ & $40.22\left[I_1 - 3\right]$ & 0.0040 \\ \hline
\multicolumn{1}{|l|}{Mooney-Rivlin\vphantom{$\frac{\int}{a}$}} & Truth & $40.00\left[I_1 - 3\right] + 20.00\left[I_2 - 3\right]$ & - \\ \cline{2-4} 
\multicolumn{1}{|l|}{\vphantom{$\frac{\int}{a}$}}            & $\sigma=0$ & $40.00\left[I_1 - 3\right] + 20.00\left[I_2 - 3\right]$ & $3.76 \cdot 10^{-32}$ \\ \cline{2-4} 
\multicolumn{1}{|l|}{\vphantom{$\frac{\int}{a}$}}            & $\sigma=5$ & $46.90\left[I_1 - 3\right] + 12.94\left[I_2 - 3\right]$ & 0.0018 \\ \hline
\multicolumn{1}{|l|}{Yeoh\vphantom{$\frac{\int}{a}$}} & Truth & $40.00\left[I_1 - 3\right] + 10.00\left[I_1 - 3\right]^2 + 30.00\left[I_1 - 3\right]^3$ & - \\ \cline{2-4} 
\multicolumn{1}{|l|}{\vphantom{$\frac{\int}{a}$}}            & $\sigma=0$ & $40.00\left[I_1 - 3\right] + 10.00\left[I_1 - 3\right]^2 + 30.00\left[I_1 - 3\right]^3$ & $3.84 \cdot 10^{-32}$ \\ \cline{2-4} 
\multicolumn{1}{|l|}{\vphantom{$\frac{\int}{a}$}}            & $\sigma=5$ & $33.76\left[I_1 - 3\right] + 40.77\left[I_1 - 3\right]^2$ & 0.0020 \\ \hline
\multicolumn{1}{|l|}{Biderman} & Truth & $40.00\left[I_1 - 3\right] + 20.00\left[I_2 - 3\right] + 10.00\left[I_1 - 3\right]^2 + 30.00\left[I_1 - 3\right]^3$ & - \\ \cline{2-4} 
\multicolumn{1}{|l|}{\vphantom{$\frac{\int}{a}$}}            & $\sigma=0$ & $55.62\left[I_1 - 3\right] + 20.21\left[I_1 - 3\right]^2 + 12.92\left[I_1-3\right] \left[I_2-3\right]$ & $1.98 \cdot 10^{-4}$ \\ \cline{2-4} 
\multicolumn{1}{|l|}{\vphantom{$\frac{\int}{a}$}}            & $\sigma=5$ & $53.14\left[I_1 - 3\right] + 14.41\left[I_1 - 3\right]^2 + 27.80\left[I_2-3\right]^2$ & 0.0011 \\ \hline
\end{tabular}
\end{table}

After generating all benchmark datasets, we apply CD for solving \refp{prob:1} assuming the linear material model library defined in \cref{sec:material_model_library_linear} considering Mooney-Rivlin features up to a polynomial order of four.
Due to the $L_1$-norm regularization, the CD algorithm results in sparse material parameter vectors and thus concise mathematical expressions for the strain energy density.
The $L_1$-norm regularization not only drives certain parameters to zero, but also induces shrinkage in the remaining nonzero parameters.
Therefore, in a postprocessing step, we use the features identified as active by the CD algorithm to solve an unregularized regression problem, i.e., \refp{prob:1} considering only the active features and $\alpha = 0$. This further decreases the model-data-mismatch while leaving the material model unchanged.

\cref{tab:benchmarks_linear} shows the material models discovered through the CD algorithm after the postprocessing step.
For all benchmarks except for the Biderman model, the correct model is discovered in the noise-free case.
For the Neo-Hookean and Mooney-Rivlin models, the correct model is also discovered for the noisy data.
For all other cases, surrogate models with satisfactory fitting accuracy and sparsity are discovered.

An in-depth discussion is required for the noise-free Biderman benchmark.
Two potential factors may prevent the exact model from being recovered.
First, for the considered experimental setup, there may exist multiple material models in the model library that exhibit identical or nearly identical stress responses.
This is evidenced by the fact that the discovered material models -- despite differing from the ground truth models -- show excellent agreement with the data.
And second, solving the $L_1$-regularized problem does not guarantee finding the best material model in the model library.
This is due to the approximation of the $L_0$-pseudo norm by the $L_1$-regularization term.


\CHANGE{We note that ISTA can readily be applied to \refp{prob:1}. However, when $f(\bfw)$ is quadratic, the CD algorithm provides a more efficient solution strategy, as it directly exploits the closed-form solution of the one-dimensional subproblems along each coordinate. Moreover, CD does not require the selection of a step-size parameter. Our numerical tests show that, in general, CD reaches the same convergence criterion approximately two orders of magnitude faster than ISTA when applied to the same optimization problem. We therefore find that CD should be preferred over ISTA in cases where $f(\bfw)$ is quadratic.}

One deficiency of using CD for solving \refp{prob:1}, is that it is not known a priori how the regularization parameter $\alpha$ must be chosen to obtain a sparse material model with low model-data-mismatch.
A suitable choice for the regularization parameter can for example be found through a manual trial-and-error procedure.
On the contrary, previous works \citep{flaschel_automated_2023-2} have proposed automated strategies for choosing the value $\alpha$ in \refp{prob:1}.
These selection strategies require to repeatedly solve the problem for different values of $\alpha$.
In this work, we investigate a third option, and apply \LARSLASSO{} for computing the regularization path of \refp{prob:1}.
We will demonstrate that knowing the regularization path simplifies the selection of a suitable value of $\alpha$.

\begin{figure}[h!]
\centering
\includegraphics[width=0.5\linewidth]{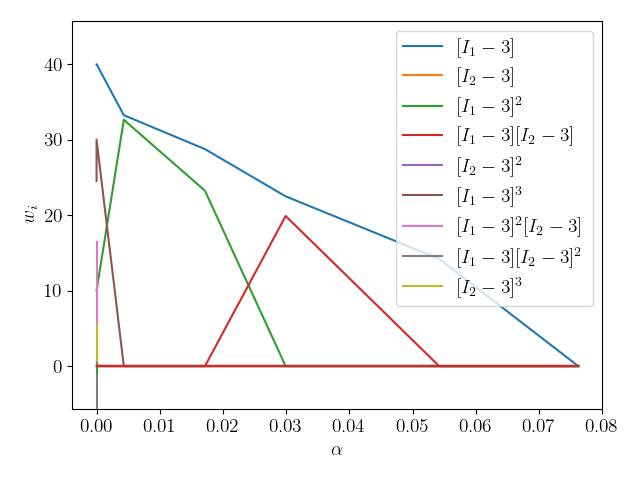}
\caption{Regularization path computed by \LARSLASSO{} for the noise-free Yeoh dataset. For clarity, legend entries of higher order features are omitted.}
\label{fig:LARS_Yeoh_noise0}
\end{figure}

\begin{table}[h!]
\caption{First steps of \LARSLASSO{} for the noise-free Yeoh dataset.}
\label{tab:LARS_Yeoh_noise0}
\centering
\begin{tabular}{|c|l|l|l|}
\hline
\multicolumn{1}{|l|}{Step\vphantom{$\frac{\int}{a}$}} & \multicolumn{1}{c|}{Strain energy density $\tilde{W}$} & \multicolumn{1}{c|}{$\alpha$} & \multicolumn{1}{c|}{$f(\bfw)$} \\ \hline
0\vphantom{$\frac{\int}{a}$} & $0.00$ & $7.62 \cdot 10^{-2}$ & $1.21 \cdot 10^{-1}$ \\ \hline
1\vphantom{$\frac{\int}{a}$} & $14.21\left[I_1 - 3\right]$ & $5.42 \cdot 10^{-2}$ & $6.31 \cdot 10^{-2}$ \\ \hline
2\vphantom{$\frac{\int}{a}$} & $22.49\left[I_1 - 3\right] + 19.88\left[I_1 - 3\right]\left[I_2 - 3\right]$ & $2.99 \cdot 10^{-2}$ & $1.99 \cdot 10^{-2}$ \\ \hline
3\vphantom{$\frac{\int}{a}$} & $28.74\left[I_1 - 3\right] + 23.24\left[I_1 - 3\right]^2$ & $1.72 \cdot 10^{-2}$ & $6.60 \cdot 10^{-3}$ \\ \hline
4\vphantom{$\frac{\int}{a}$} & $33.24\left[I_1 - 3\right] + 32.63\left[I_1 - 3\right]^2$ & $2.32 \cdot 10^{-3}$ & $6.14 \cdot 10^{-4}$ \\ \hline
5\vphantom{$\frac{\int}{a}$} & $40.00\left[I_1 - 3\right] + 10.00\left[I_1 - 3\right]^2 + 30.00\left[I_1 - 3\right]^3$ & $5.43 \cdot 10^{-16}$ & $9.21 \cdot 10^{-30}$ \\ \hline
\end{tabular}
\end{table}

To demonstrate the functionality of \LARSLASSO{}, we show the computed regularization path in \cref{fig:LARS_Yeoh_noise0} as well as the first steps of \LARSLASSO{} in \cref{tab:LARS_Yeoh_noise0}, both for the noise-free Yeoh dataset.
It is observed that after five iterations, the model-data-mismatch decreases to effectively zero, as \LARSLASSO{} has discovered the ground truth material model by identifying the correct modeling features.

Notably, \LARSLASSO{} does not only consider correct features during the first iterations.
After the second iteration, \LARSLASSO{} identifies a false-positive feature, i.e., a feature that appears in the discovered model while not appearing in the Yeoh model.
However, after the third iteration, this feature is eliminated.
This behavior can be traced back to the modification applied to \LARS{} in \cref{sec:LARSLASSO}.

At this point, it is important to mention that \LARSLASSO{} computes the knots of the regularization path.
Consequently, not each step of \LARSLASSO{} corresponds to a critical value $\alpha_c$ as defined in \refp{prob:2}, see \cref{fig:regularization_path_knots_L0}.
However, for a given set of knots, the critical values can be easily extracted.
For example, in the second to fourth steps in \cref{tab:LARS_Yeoh_noise0}, the parameter vectors exhibit the same number of nonzero parameters.
Therefore, out of these three steps, only the fourth step corresponds to a critical value.
Furthermore, as only the initial steps of \LARSLASSO{} are of interest in practice, and due to the early stopping criterion described in \cref{sec:early_stopping}, \LARSLASSO{} does not identify all critical values, but only the practically relevant initial ones.

\cref{tab:LARS_Yeoh_noise0} highlights the important advantages of \LARSLASSO{} over methods like CD for solving \refp{prob:1}.
When approaching \refp{prob:1} with the CD algorithm, a suitable value for $\alpha$ must be found either through trial-and-error or by solving the problem for multiple values of $\alpha$ and conducting a Pareto analysis \citep{flaschel_automated_2023-2}.
\LARSLASSO{} always starts with the smallest value of $\alpha$ for which all parameters are zero and then subsequently decreases $\alpha$.
Importantly, $\alpha$ is not decreased in equidistant steps, but with varying step sizes.
The algorithm guarantees that no significant changes occur between two consecutive steps, i.e., between two steps no features are added or removed from the active set.
In this way, \LARSLASSO{} efficiently identifies the critical and practically meaningful values for $\alpha$.
For example, between the fourth and fifth steps, $\alpha$ decreases by several orders of magnitude, as \LARSLASSO{} automatically identifies that no significant changes occur across these orders of magnitude.
Choosing values of $\alpha$ that are between two \LARSLASSO{} steps is practically not meaningful, as there exists a smaller value of $\alpha$ yielding the same material model.
\LARSLASSO{} thus identifies intervals of $\alpha$ that have no impact on the model and can be disregarded.

\begin{figure}[h!]
\centering
\includegraphics[width=0.5\linewidth]{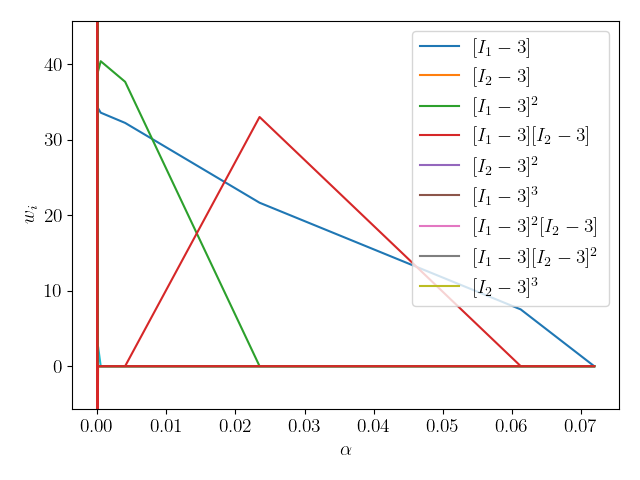}
\caption{Regularization path computed by \LARSLASSO{} for the noisy Yeoh dataset. For clarity, legend entries of higher order features are omitted.}
\label{fig:LARS_Yeoh_noise5}
\end{figure}

\begin{table}[h!]
\caption{First steps of \LARSLASSO{} for the noisy Yeoh dataset.}
\label{tab:LARS_Yeoh_noise5}
\centering
\begin{tabular}{|c|l|l|l|}
\hline
\multicolumn{1}{|l|}{Step\vphantom{$\frac{\int}{a}$}} & \multicolumn{1}{c|}{Strain energy density $\tilde{W}$} & \multicolumn{1}{c|}{$\alpha$} & \multicolumn{1}{c|}{$f(\bfw)$} \\ \hline
0\vphantom{$\frac{\int}{a}$} & $0.00$ & $7.19 \cdot 10^{-2}$ & $0.1102$ \\ \hline
1\vphantom{$\frac{\int}{a}$} & $7.52\left[I_1 - 3\right]$ & $6.13 \cdot 10^{-2}$ & $0.0819$ \\ \hline
2\vphantom{$\frac{\int}{a}$} & $21.65\left[I_1 - 3\right] + 33.00\left[I_1 - 3\right]\left[I_2 - 3\right]$ & $2.35 \cdot 10^{-2}$ & $0.0142$ \\ \hline
3\vphantom{$\frac{\int}{a}$} & $32.21\left[I_1 - 3\right] + 37.66\left[I_1 - 3\right]^2$ & $4.03 \cdot 10^{-3}$ & $0.0023$ \\ \hline
4\vphantom{$\frac{\int}{a}$} & $33.57\left[I_1 - 3\right] + 40.39\left[I_1 - 3\right]^2$ & $4.91 \cdot 10^{-4}$ & $0.0020$ \\ \hline
5\vphantom{$\frac{\int}{a}$} & $34.15\left[I_1 - 3\right] + 39.07\left[I_1 - 3\right]^2 + 2.58\left[I_1 - 3\right]^4$ & $1.16 \cdot 10^{-4}$ & $0.0020$ \\ \hline
\end{tabular}
\end{table}

Next, we consider the noisy dataset corresponding to the Yeoh model. \cref{fig:LARS_Yeoh_noise5} and \cref{tab:LARS_Yeoh_noise5} show the regularization path identified by \LARSLASSO{}. After the fourth step of \LARSLASSO{}, the model-data-mismatch is barely decreasing.
Therefore, $\alpha = 4.91 \cdot 10^{-4}$ is a good choice for this example.
Notably, as a result of the noise, \LARSLASSO{} does not discover the Yeoh model, but a surrogate model that fits the dataset while being expressed as a short mathematical expression.

\begin{figure}[h!]
\centering
\begin{subfigure}{0.4\textwidth}
\includegraphics[width=\linewidth]{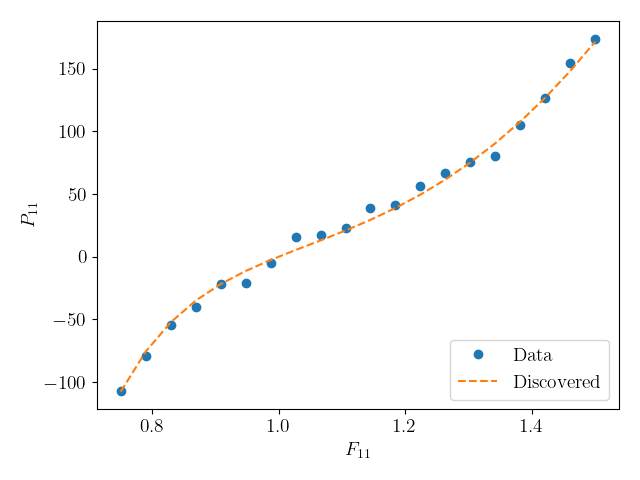}
\caption{Uniaxial tension and compression.}
\end{subfigure}%
\hspace{1.5cm}
\begin{subfigure}{0.4\textwidth}
\includegraphics[width=\linewidth]{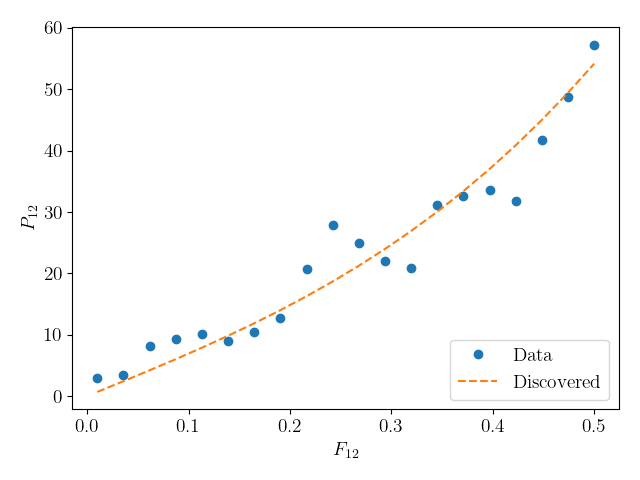}
\caption{Simple shear.}
\end{subfigure}%
\vspace{0.0cm}
\caption{Stress-strain response of the discovered material model for the noisy Yeoh dataset.
}
\label{fig:LARS_Yeoh_noise5_UCTSS}
\end{figure}

We apply \LARSLASSO{} in the same manner to all datasets.
For each benchmark, we choose a \LARSLASSO{} step at which the solution is sparse and the model-data-mismatch is low, and afterwards apply the previously described postprocessing step to further reduce the model-data-mismatch.
The discovered material models are equal to those discovered with the CD algorithm, see \cref{tab:benchmarks_linear}.

\subsection{Nonlinear material model library}

Next, we consider benchmark problems with material models that depend nonlinearly on the parameters, and apply ISTA for material model discovery.
Specifically, we choose the Mooney-Rivlin model, the Ogden model and a mixed model that contains both the Mooney-Rivlin and Ogden features, see \cref{tab:benchmarks_nonlinear}.
We generate synthetic data in a way analogous to \cref{sec:results_linear} and perturb them by independent Gaussian noise.

\begin{table}[h!]
\caption{Strain energy density functions of the true and discovered material models and model-data-mismatch.}
\label{tab:benchmarks_nonlinear}
\centering
\begin{tabular}{|ll|l|r|}
\hline
\multicolumn{2}{|l|}{Benchmarks\vphantom{$\frac{\int}{\int}$}} & \multicolumn{1}{c|}{Strain energy density $\tilde{W}$} & \multicolumn{1}{c|}{$f(\bfw)$} \\ \hline
\multicolumn{1}{|l|}{Mooney-Rivlin\vphantom{$\frac{\int}{\int}$}} & Truth & $40.00\left[I_1 - 3\right] + 20.00\left[I_2 - 3\right]$ & - \\ \cline{2-4} 
\multicolumn{1}{|l|}{\vphantom{$\frac{\int}{\int}$}}            & $\sigma=0$ & $21.07\left[I_2 - 3\right] + 24.61[\lambda_1^{2.48} + \lambda_2^{2.48} + \lambda_3^{2.48} - 3]$ & $4.28 \cdot 10^{-05}$ \\ \cline{2-4} 
\multicolumn{1}{|l|}{\vphantom{$\frac{\int}{\int}$}}            & $\sigma=5$ & $17.22\left[I_2 - 3\right] + 23.74[\lambda_1^{2.64} + \lambda_2^{2.64} + \lambda_3^{2.64} - 3]$ & 0.0019 \\ \hline
\multicolumn{1}{|l|}{Ogden\vphantom{$\frac{\int}{\int}$}} & Truth & $5.00[\lambda_1^{8.00} + \lambda_2^{8.00} + \lambda_3^{8.00} - 3]$ & - \\ \cline{2-4} 
\multicolumn{1}{|l|}{\vphantom{$\frac{\int}{\int}$}}            & $\sigma=0$ & $4.94[\lambda_1^{8.03} + \lambda_2^{8.03} + \lambda_3^{8.03} - 3]$ & $6.21 \cdot 10^{-07}$ \\ \cline{2-4} 
\multicolumn{1}{|l|}{\vphantom{$\frac{\int}{\int}$}}            & $\sigma=5$ & $4.99[\lambda_1^{8.04} + \lambda_2^{8.04} + \lambda_3^{8.04} - 3]$ & 0.0003 \\ \hline
\multicolumn{1}{|l|}{Mixed Model\vphantom{$\frac{\int}{\int}$}} & Truth & $40.00\left[I_1 - 3\right] + 20.00\left[I_2 - 3\right] + 5.00[\lambda_1^{8.00} + \lambda_2^{8.00} + \lambda_3^{8.00} - 3]$ & - \\ \cline{2-4} 
\multicolumn{1}{|l|}{\vphantom{$\frac{\int}{\int}$}}            & $\sigma=0$ & $12.85[\lambda_1^{6.54} + \lambda_2^{6.54} + \lambda_3^{6.54} - 3]$ & $5.57 \cdot 10^{-05}$ \\ \cline{2-4} 
\multicolumn{1}{|l|}{\vphantom{$\frac{\int}{\int}$}}            & $\sigma=5$ & $13.19[\lambda_1^{6.51} + \lambda_2^{6.51} + \lambda_3^{6.51} - 3]$ & 0.0005 \\ \hline
%
\end{tabular}
\end{table}

We apply ISTA assuming the nonlinear material model library defined in \cref{sec:material_model_library_nonlinear}.
For all benchmarks, we choose $\bfw^{(0)} = \boldsymbol{1}$ as the initial guess to demonstrate the sparsity-promoting property of the $L_1$-regularization term. 
As described earlier, after ISTA has converged and a material model with several vanishing material parameters has been discovered, as a postprocessing step, we solve the unregularized \refp{prob:3} while keeping the zero parameters fixed.
Again, this is motivated in further reducing the model-data-mismatch while the model remains unchanged.
\cref{tab:benchmarks_nonlinear} shows the material models discovered for different noise levels.
For the Ogden model, ISTA discovers the correct material models both in the noise-free and the noisy cases, and the discovered models exhibit a small model-data-mismatch.
For the Mooney-Rivlin model and the mixed model, however, despite showing a low model-data-mismatch, the discovered material models do not match the ground truth models.
As discussed previously, this can be a result of the approximation of the $L_0$-pseudo norm by the $L_1$-regularization term, which does not guarantee that the best model in the library is found.
Additionally, there may be similar features in the model library, which makes it difficult to recover the exact material model in the inverse problem.
In fact, the first Mooney-Rivlin feature is equal to the Ogden feature for $\delta = 2$, i.e., $I_1 = \lambda_1^{2} + \lambda_2^{2} + \lambda_3^{2}$.
This could explain why ISTA discovers the Ogden feature instead of the first Mooney-Rivlin feature.
In practical applications, the ground truth model is generally unavailable.
Therefore, the primary objective is typically to identify a sparse model that exhibits minimal discrepancy between the model and the observed data, an objective that ISTA is well-suited to achieve.

\begin{figure}[h!]
\centering
\begin{subfigure}{0.33\textwidth}
\includegraphics[width=\linewidth]{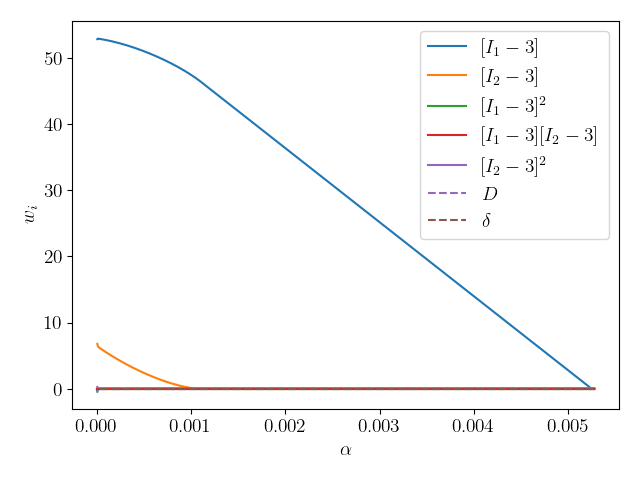}
\caption{Mooney-Rivlin.}
\end{subfigure}%
\hspace{0.0cm}
\begin{subfigure}{0.33\textwidth}
\includegraphics[width=\linewidth]{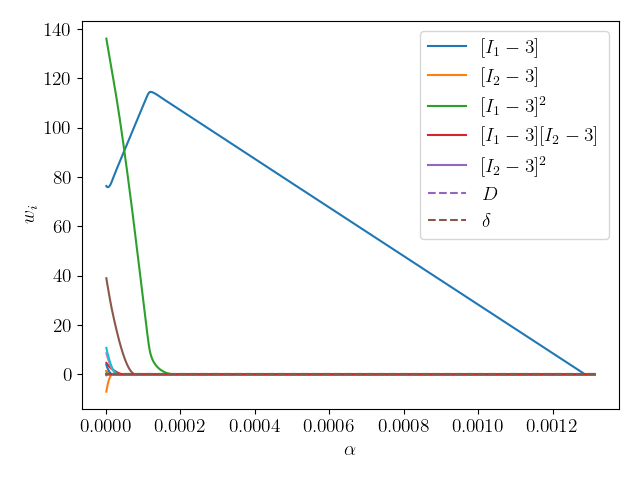}
\caption{Ogden.}
\end{subfigure}%
\hspace{0.0cm}
\begin{subfigure}{0.33\textwidth}
\includegraphics[width=\linewidth]{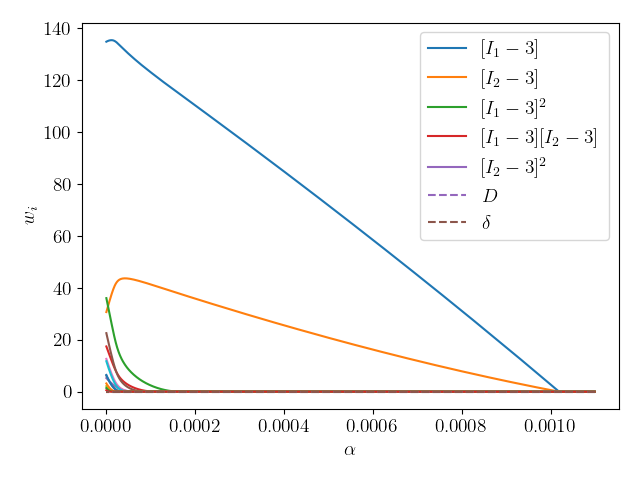}
\caption{Mixed model.}
\end{subfigure}%
\vspace{0.0cm}
\caption{Regularization path computed by the pathwise ISTA for the noisy datasets. For clarity, legend entries of higher order Mooney-Rivlin features are omitted.
}
\label{fig:pathwise_ISTA}
\end{figure}

Finally, we use the pathwise ISTA with $n_\alpha = 1000$ to approximately compute the regularization path.
\cref{fig:pathwise_ISTA} shows the results for the noisy datasets.
As $\alpha$ decreases, the number of nonzero material parameters and the fitting accuracy of the models increase.
The pathwise ISTA correctly identifies the Mooney-Rivlin features for the Mooney-Rivlin model and the mixed model.
However, for the reasons discussed above, it fails to identify the Ogden feature in both the Ogden and mixed models, instead selecting surrogate features that mimic the behavior of the true Ogden feature.
We note that, due to the different initial guesses, the regularization path may not contain the models discovered by ISTA shown in \cref{tab:benchmarks_nonlinear}.
Specifically, for the results in \cref{tab:benchmarks_nonlinear}, we assumed the initial guess $\bfw^{(0)} = \boldsymbol{1}$, while the initial guess in each step of the pathwise ISTA depends on the result of the previous step.

\section{Conclusions and outlook}
\label{sec:conclusions}

The field of non-smooth optimization provides a range of tools for addressing challenges in material model discovery.
In this work, we have discovered material models that depend linearly or nonlinearly on the material parameters using the CD algorithm and ISTA, respectively.
These methods robustly solve the underlying sparse regression problems with proven convergence for a given value of the regularization parameter.
\CHANGE{
Using the CD algorithm is preferred over ISTA when the material model depends linearly on the material parameters, as CD does not require step size selection and efficiently leverages the closed-form solution of the corresponding one-parameter problem.
}
For material models that depend nonlinearly on the material parameters, however, ISTA offers an attractive alternative with proven convergence if the step size is sufficiently small.
Both the CD algorithm and ISTA require choosing the regularization parameter a priori.
Conversely, the \LARSLASSO{} algorithm leverages the piecewise linearity of the regularization path to efficiently compute the critical values of the regularization parameter at which the number of nonzero elements in the material parameter vector changes.
This facilitates the manual selection of a material model from the first steps of \LARSLASSO{}.
For mechanics applications, where we are typically interested in material models with only a few nonzero parameters, \LARSLASSO{} offers an efficient alternative to the CD algorithm -- especially when it is terminated after the first few iterations to avoid computing the full regularization path.
\CHANGE{
Finally, by successively decreasing the regularization parameter and using the previous solutions as initial guesses for the subsequent solves, the novel pathwise ISTA efficiently computes the nonlinear regularization path when the material models depend nonlinearly on the material parameters.
}
The concepts presented in this work can be extended in several directions.
For example, we did not consider constraints on the material parameters.
Such constraints can be incorporated into the CD algorithm, \LARSLASSO{}, and ISTA without significant effort.
Additionally, for material models that depend nonlinearly on the material parameters, we have focused on the first-order method ISTA.
In the future, second-order methods that exploit the second derivative of the model-data mismatch may be explored for material model discovery.
In the future, we aim to investigate all algorithms discussed in this paper in the context of dissipative materials and to apply the discussed algorithms to experimentally measured data.

\section*{Code and data availability}

Code and data are publicly available on Zenodo (\url{https://doi.org/10.5281/zenodo.15848305}), see \cite{flaschel_supplementary_2025}, and on GitHub at \url{https://github.com/mflaschel/non-smooth-material-model-discovery}.

\section*{Acknowledgments}

Moritz Flaschel thanks Clemens Sirotenko for fruitful discussions and literature suggestions.
This work was supported by the ERC Advanced Grant 101141626 DISCOVER to Ellen Kuhl.
The authors utilized HAWKI, a large language model interface provided by FAU, to enhance the writing style in certain sections of the manuscript. After using this tool, the authors reviewed and edited the content as needed and take full responsibility for the content of the publication.


\appendix




\section{Boundedness of the regularization parameter}
\label{sec:boundedness_regularization_parameter}
Choosing a large value of $\alpha$ in \refp{prob:1} or \refp{prob:3} may result in a solution vector whose entries are all zero, i.e., $\bfw^*=\bf0$.
As we are typically not interested in this zero solution, we seek to choose smaller values of $\alpha$ in practice.
We are thus interested in determining the minimum value of $\alpha$ such that $\bfw^*=\bf0$, which we will denote by $\alpha^{(0)}$.
To this end, we consider the necessary condition for a minimum of \refp{prob:3}
\be
\label{eq:necessary_condition_boundedness_regularization_parameter}
0 \in \frac{\partial f}{\partial w_i}(\boldsymbol{0}) + \alpha^{(0)} [-1,1],
\ee
which must hold true for all $i$.
After shifting the partial derivative of $f$ to the left and taking the absolute value, we obtain
\be
\left|\frac{\partial f}{\partial w_i}(\boldsymbol{0})\right| \in \alpha^{(0)} [0,1].
\ee
The minimum value of $\alpha$ that fulfills this condition for all $i$ is
\be
\alpha^{(0)} = \max_i \left|\frac{\partial f}{\partial w_i}(\boldsymbol{0})\right|.
\ee
In practice, we are interested in values of $\alpha$ that do not yield the zero solution, i.e.,
\be
\alpha \in (0,\alpha^{(0)}).
\ee
We note that \cref{eq:necessary_condition_boundedness_regularization_parameter} is a necessary but not sufficient condition. Thus, dependent on the appearance of $f$, $\alpha^{(0)}$ may not necessarily yield the zero solution.
Nevertheless, we will adhere to the above range of $\alpha$ in this work.

Finally, for the special case in \refp{prob:1}, we obtain \citep{kim_interior-point_2007}
\be
\alpha^{(0)} = \frac{1}{n} \max_i | \bfX_i^T \bfy |.
\ee

\section{Additional information to \LARS{} and \LARSLASSO{}}

\subsection{Boundedness of the step size}
\label{sec:bounded_step_size}


It can be shown that the step size of \LARS{}, see \cref{eq:LARS_stepsize}, is bounded by $0 < \gamma^{(k)} < \frac{\bar{c}^{(k)}_{\text{max}}}{A^{(k)}}$ \citep{efron_least_2004}.
To show it, we define
\be
\label{eq:LARS_stepsize}
\gamma^{(k)}_{j^*} = {\min}^+ \left\{
\frac{\bar{c}^{(k)}_{\text{max}} - c^{(k)}_{j^*}}{A^{(k)} - a^{(k)}_{j^*}},
\frac{\bar{c}^{(k)}_{\text{max}} + c^{(k)}_{j^*}}{A^{(k)} + a^{(k)}_{j^*}}
\right\},
\ee
and show that $0 < \gamma^{(k)}_{j^*} < \frac{\bar{c}^{(k)}_{\text{max}}}{A^{(k)}}$ for all $j^* \in \mathcal{A}^{\complement(k)}$.
First, we observe that $\bar{c}^{(k)}_{\text{max}} - c^{(k)}_{j^*} > 0$ and $\bar{c}^{(k)}_{\text{max}} + c^{(k)}_{j^*} > 0$, and thus $\gamma^{(k)}_{j^*} > 0$. Next, we distinguish the cases $a^{(k)}_{j^*} < 0$ and $a^{(k)}_{j^*}>0$, where the case $a^{(k)}_{j^*} = 0$ is apparent.
\begin{itemize}
    \item Case 1: We consider the case $a^{(k)}_{j^*} < 0$.
    We further distinguish the sub-cases $A^{(k)} + a^{(k)}_{j^*} < 0$ and $A^{(k)} + a^{(k)}_{j^*} > 0$.
    \begin{itemize}
        \item Sub-case 1.1: We consider the sub-case $A^{(k)} + a^{(k)}_{j^*} < 0$.
        It follows $\frac{\bar{c}^{(k)}_{\text{max}} + c^{(k)}_{j^*}}{A^{(k)} + a^{(k)}_{j^*}} < 0$.
        Thus, $\gamma^{(k)}_{j^*} = \frac{\bar{c}^{(k)}_{\text{max}} - c^{(k)}_{j^*}}{A^{(k)} - a^{(k)}_{j^*}} < \frac{\bar{c}^{(k)}_{\text{max}}}{A^{(k)}}$.
        \item Sub-case 1.2: We consider the sub-case $A^{(k)} + a^{(k)}_{j^*} > 0$.
        It follows $0 < \frac{\bar{c}^{(k)}_{\text{max}} - c^{(k)}_{j^*}}{A^{(k)} - a^{(k)}_{j^*}} < \frac{\bar{c}^{(k)}_{\text{max}} + c^{(k)}_{j^*}}{A^{(k)} + a^{(k)}_{j^*}}$.
        Thus, $\gamma^{(k)}_{j^*} = \frac{\bar{c}^{(k)}_{\text{max}} - c^{(k)}_{j^*}}{A^{(k)} - a^{(k)}_{j^*}} < \frac{\bar{c}^{(k)}_{\text{max}}}{A^{(k)}}$.
    \end{itemize}

    \item Case 2: We consider the case $a^{(k)}_{j^*}>0$.
    We further distinguish the sub-cases $A^{(k)} - a^{(k)}_{j^*} < 0$ and $A^{(k)} - a^{(k)}_{j^*} > 0$.
    \begin{itemize}
        \item Sub-case 2.1: We consider the sub-case $A^{(k)} - a^{(k)}_{j^*} < 0$.
        It follows $\frac{\bar{c}^{(k)}_{\text{max}} - c^{(k)}_{j^*}}{A^{(k)} - a^{(k)}_{j^*}} < 0$.
        Thus, $\gamma^{(k)}_{j^*} = \frac{\bar{c}^{(k)}_{\text{max}} + c^{(k)}_{j^*}}{A^{(k)} + a^{(k)}_{j^*}}$.
        We notice that $\bar{c}^{(k)}_{\text{max}} + c^{(k)}_{j^*} < 2 \bar{c}^{(k)}_{\text{max}}$ and $A^{(k)} + a^{(k)}_{j^*} > 2 A^{(k)}$. Therefore, $\gamma^{(k)}_{j^*} <  \frac{\bar{c}^{(k)}_{\text{max}}}{A^{(k)}}$.
        \item Sub-case 2.2: We consider the sub-case $A^{(k)} - a^{(k)}_{j^*} > 0$.
        We further distinguish the sub-sub-cases $a^{(k)}_{j^*} < \frac{A^{(k)}}{\bar{c}^{(k)}_{\text{max}}}c^{(k)}_{j^*}$ and $a^{(k)}_{j^*} > \frac{A^{(k)}}{\bar{c}^{(k)}_{\text{max}}}c^{(k)}_{j^*}$. 
        For $a^{(k)}_{j^*} < \frac{A^{(k)}}{\bar{c}^{(k)}_{\text{max}}}c^{(k)}_{j^*}$, it is $0 < \frac{\bar{c}^{(k)}_{\text{max}} - c^{(k)}_{j^*}}{A^{(k)} - a^{(k)}_{j^*}} < \frac{\bar{c}^{(k)}_{\text{max}}}{A^{(k)}}$, and for $a^{(k)}_{j^*} > \frac{A^{(k)}}{\bar{c}^{(k)}_{\text{max}}}c^{(k)}_{j^*}$, it is $0 < \frac{\bar{c}^{(k)}_{\text{max}} + c^{(k)}_{j^*}}{A^{(k)} + a^{(k)}_{j^*}} < \frac{\bar{c}^{(k)}_{\text{max}}}{A^{(k)}}$. Therefore, $\gamma^{(k)}_{j^*} <  \frac{\bar{c}^{(k)}_{\text{max}}}{A^{(k)}}$.
    \end{itemize}
\end{itemize}


\subsection{How inactive parameters enter the active set}
\label{sec:enter_the_active_set}

Inactive parameters can be shown to enter the active set such that their sign is equal to the sign of the correlation of the corresponding feature vector \citep{efron_least_2004}.
Formally, this can be formulated as follows.
We consider a step $k$ in which one parameter $w_l$ enters the active set, i.e., $w^{(k)}_l = 0$ and $w^{(k+1)}_l \neq 0$, with $l$ such that $l \notin \mathcal{A}^{(k-1)}$ and $l \in \mathcal{A}^{(k)} = \mathcal{A}^{(k-1)} \cup \{ l \}$.
It can be shown that $\sign(w^{(k+1)}_l) = \sign(c^{(k)}_l)$.

We first consider the first step $k=0$ for which the active set reduces to $\mathcal{A}^{(0)} = \{ l \}$.
It is $w^{(1)}_l = \gamma^{(0)} \bar{c}^{(0)}_u \sign(c^{(0)}_l)$, where we used $w^{(0)}_l = 0$ and $\bar{X}^{T(0)}_{l}\bar{X}^{(0)}_{l} = 1$. Because $\gamma^{(0)}>0$, $\bar{c}^{(0)}_u>0$, it is $\sign(w^{(1)}_l) = \sign(c^{(0)}_l)$.

Next, we focus on the general case $k>1$.
We recall that $\bfw^{(k+1)}_{\mathcal{A}} = \bfw^{(k)}_{\mathcal{A}} + \Delta\bfw^{(k)}_{\mathcal{A}}$, where we have $w^{(k)}_l = 0$ for the element of interest, such that $w^{(k+1)}_l = \Delta w^{(k)}_l$.
The parameter updates are computed according to 
\be
\Delta\bfw^{(k)}_{\mathcal{A}} = \gamma^{(k)} A^{(k)} \diag(\bfs^{(k)}_{\mathcal{A}})\left[\bar{\bfX}^{T(k)}_{\mathcal{A}}\bar{\bfX}^{(k)}_{\mathcal{A}}\right]^{-1}\boldsymbol{1}_{\mathcal{A}}.
\ee
Because all feature vectors of the active set share the same correlation in absolute value, we may substitute
\be
\boldsymbol{1}_{\mathcal{A}} = \frac{\bar{c}^{(k)}_{\text{max}}}{\bar{c}^{(k)}_{\text{max}}} \boldsymbol{1}_{\mathcal{A}} = \frac{1}{\bar{c}^{(k)}_{\text{max}}} \bar{\bfX}^{T(k)}_{\mathcal{A}} [\bfy_{\parallel} - \bfmu^{(k)}],
\ee
to obtain
\be
\Delta\bfw^{(k)}_{\mathcal{A}} = \gamma^{(k)} \frac{A^{(k)}}{\bar{c}^{(k)}_{\text{max}}} \diag(\bfs^{(k)}_{\mathcal{A}})
\underbrace{\left[\bar{\bfX}^{T(k)}_{\mathcal{A}}\bar{\bfX}^{(k)}_{\mathcal{A}}\right]^{-1}\bar{\bfX}^{T(k)}_{\mathcal{A}} [\bfy_{\parallel} - \bfmu^{(k)}]}_{\bfw^{(k)}_{\parallel}},
\ee
where we identify $\bfw^{(k)}_{\parallel}$ as the least squares solution $\bfw^{(k)}_{\parallel} = \argmin_{\bfw} \| \bar{\bfX}^{(k)}_{\mathcal{A}} \bfw - \bfr^{(k)}_{\parallel} \|^2$.
It is $\Delta w^{(k)}_l = \gamma^{(k)} \frac{A^{(k)}}{\bar{c}^{(k)}_{\text{max}}} \sign(c^{(k)}_l) w^{(k)}_{\parallel \ l}$, and therefore $\sign(w^{(k+1)}_l) = \sign(c^{(k)}_l)$ if $w^{(k)}_{\parallel \ l} > 0$.

Thus, the remaining task is to show that $w^{(k)}_{\parallel \ l}$ is positive.
To this end, we additively divide $\bfy_{\parallel}$ into the two contributions $\bfy^{(k-1)}_{\parallel}$ and $\bfy_{\parallel}-\bfy^{(k-1)}_{\parallel}$.
Specifically, we define $\bfy^{(k-1)}_{\parallel} = \bar{\bfX}^{(k-1)}_{\mathcal{A}}\left[\bar{\bfX}^{T(k-1)}_{\mathcal{A}}\bar{\bfX}^{(k-1)}_{\mathcal{A}}\right]^{-1}\bar{\bfX}^{T(k-1)}_{\mathcal{A}}\bfy$, which we may write as
\be
\begin{aligned}
\bfy^{(k-1)}_{\parallel}
&= \bar{\bfX}^{(k-1)}_{\mathcal{A}}\left[\bar{\bfX}^{T(k-1)}_{\mathcal{A}}\bar{\bfX}^{(k-1)}_{\mathcal{A}}\right]^{-1}\bar{\bfX}^{T(k-1)}_{\mathcal{A}}[\bfy_{\parallel} + \bfy_{\perp} + \bfmu^{(k-1)} - \bfmu^{(k-1)}] \\
&= \bfmu^{(k-1)} + \bar{\bfX}^{(k-1)}_{\mathcal{A}}\left[\bar{\bfX}^{T(k-1)}_{\mathcal{A}}\bar{\bfX}^{(k-1)}_{\mathcal{A}}\right]^{-1}\bar{\bfX}^{T(k-1)}_{\mathcal{A}}[\bfy_{\parallel} - \bfmu^{(k-1)}] \\
&= \bfmu^{(k-1)} + \bar{\bfX}^{(k-1)}_{\mathcal{A}}\left[\bar{\bfX}^{T(k-1)}_{\mathcal{A}}\bar{\bfX}^{(k-1)}_{\mathcal{A}}\right]^{-1}\bar{c}^{(k-1)}_{\text{max}}\boldsymbol{1}_{\mathcal{A}}\\
&= \bfmu^{(k-1)} + \frac{\bar{c}^{(k-1)}_{\text{max}}}{A^{(k-1)}} \bfu^{(k-1)},
\end{aligned}
\ee
where we used $\bar{\bfX}^{T(k-1)}_{\mathcal{A}}\bfy_{\perp} = \bf0$, $\bfmu^{(k-1)} = \bfX \bfw^{(k-1)} = \bar{\bfX}^{(k-1)}_{\mathcal{A}} \diag(\bfs^{(k-1)}_{\mathcal{A}}) \bfw^{(k-1)}_{\mathcal{A}}$, and $\bar{\bfX}^{T(k-1)}_{\mathcal{A}}\bfr_{\parallel}^{(k-1)} = \bar{c}^{(k-1)}_{\text{max}}\boldsymbol{1}_{\mathcal{A}}$.

Recalling that $\bfmu^{(k)}=\bfmu^{(k-1)} + \gamma^{(k-1)} \bfu^{(k-1)}$, we can rewrite $\bfw^{(k)}_{\parallel}$ as
\be
\begin{aligned}
\bfw^{(k)}_{\parallel}
&= \left[\bar{\bfX}^{T(k)}_{\mathcal{A}}\bar{\bfX}^{(k)}_{\mathcal{A}}\right]^{-1}\bar{\bfX}^{T(k)}_{\mathcal{A}} [\bfy_{\parallel} - \bfy^{(k-1)}_{\parallel} + \bfy^{(k-1)}_{\parallel} - \bfmu^{(k-1)} + \gamma^{(k-1)} \bfu^{(k-1)}] \\
&= \left[\bar{\bfX}^{T(k)}_{\mathcal{A}}\bar{\bfX}^{(k)}_{\mathcal{A}}\right]^{-1}\bar{\bfX}^{T(k)}_{\mathcal{A}} \left[\bfy_{\parallel} - \bfy^{(k-1)}_{\parallel} + \left(\frac{\bar{c}^{(k-1)}_{\text{max}}}{A^{(k)}} + \gamma^{(k-1)} \right) \bfu^{(k-1)}\right] \\
&= \underbrace{\left[\bar{\bfX}^{T(k)}_{\mathcal{A}}\bar{\bfX}^{(k)}_{\mathcal{A}}\right]^{-1}\bar{\bfX}^{T(k)}_{\mathcal{A}} [\bfy_{\parallel} - \bfy^{(k-1)}_{\parallel} ]}_{\bfw^{(k)}_{\bfy}}
+ \underbrace{\left(\frac{\bar{c}^{(k-1)}_{\text{max}}}{A^{(k)}} + \gamma^{(k-1)} \right) \left[\bar{\bfX}^{T(k)}_{\mathcal{A}}\bar{\bfX}^{(k)}_{\mathcal{A}}\right]^{-1}\bar{\bfX}^{T(k)}_{\mathcal{A}}  \bfu^{(k-1)}}_{\bfw^{(k)}_{\bfu}}.
\end{aligned}
\ee
By the definition of $\bfy^{(k-1)}_{\parallel}$, it is $\bar{\bfX}^{T(k-1)}_{\mathcal{A}}[\bfy-\bfy^{(k-1)}_{\parallel}] = \bf0$.
Consequently, we observe that $\bar{\bfX}^{T(k)}_{i} [\bfy-\bfy^{(k-1)}_{\parallel}]$ must be zero for all $i \in \mathcal{A}^{(k-1)}$ and can only be nonzero for $i=l$.
We define the vector $\bfdelta$ such that $\delta_i = 0$ if $i \neq l$ and $\delta_l = \bar{\bfX}^{T(k)}_{l} [\bfy-\bfy^{(k-1)}_{\parallel}]$.
Thus, it is $\bfw^{(k)}_{\bfy} = \left[\bar{\bfX}^{T(k)}_{\mathcal{A}}\bar{\bfX}^{(k)}_{\mathcal{A}}\right]^{-1}\bfdelta$, where the element of interest is $w^{(k)}_{\bfy \ l} = \left\{\left[\bar{\bfX}^{T(k)}_{\mathcal{A}}\bar{\bfX}^{(k)}_{\mathcal{A}}\right]^{-1}\right\}_{ll}\delta_l$, where no sum of repeated indices is applied.
We find that $w^{(k)}_{\bfy \ l}$ is positive because the diagonal elements of the positive definite matrix $\left[\bar{\bfX}^{T(k)}_{\mathcal{A}}\bar{\bfX}^{(k)}_{\mathcal{A}}\right]^{-1}$ are positive and $\delta_l$ is positive, see \cite{efron_least_2004}.
Finally, $w^{(k)}_{\bfu \ l}$ vanishes because $\bfu^{(k-1)}$ is a linear combination of the feature vectors $\bar{\bfX}^{T(k-1)}_{\mathcal{A}}$.

\subsection{Computation of the regularization parameter}
\label{sec:regularization_parameter_LARSLASSO}

The solutions $\bfw^{(k)}$ obtained by \LARSLASSO{} are solutions to \refp{prob:1} \citep{efron_least_2004} for different choices of the regularization parameter.
This raises the question of whether we can determine the regularization parameter $\alpha^{(k)}$ corresponding to a given solution $\bfw^{(k)}$.
Specifically, given that
\be
\bfw^{(k)} = \argmin_{\bfw} \frac{1}{2n} \| \bfy - \bfX \bfw \|_2^2 + \alpha^{(k)} \|\bfw\|_1,
\ee
we seek to find $\alpha^{(k)}$.
For the special case $k=0$, we refer to \cref{sec:boundedness_regularization_parameter}.
The solution $\bfw^{(k)}$ for $k>0$ must fulfill the necessary condition for a minimum
\be
\boldsymbol{0} \in - \frac{1}{n} \bfX^T ( \bfy - \bfX \bfw^{(k)} ) + \alpha^{(k)} \sign(\bfw^{(k)})
= - \frac{1}{n} \bfc^{(k)} + \alpha^{(k)} \sign(\bfw^{(k)}),
\ee
where we defined
\be
\sign(\bfw^{(k)})_i = 
\begin{cases}
    \{-1\} & \text{if } w^{(k)}_i < 0 \\
    [-1,1] & \text{if } w^{(k)}_i = 0 \\
    \{1\} & \text{if } w^{(k)}_i > 0 
\end{cases}.
\ee
Thus, for a parameter $w^{(k)}_{i^*} \neq 0$ with $i^* \in \mathcal{A}^{(k)}$, we obtain
\be
0 = - \frac{1}{n} c^{(k)}_{i^*} + \alpha^{(k)} \sign(w^{(k)}_{i^*}),
\ee
from which we deduce the regularization parameter
\be
\alpha^{(k)} = \frac{c^{(k)}_{i^*}}{\sign(w^{(k)}_{i^*}) n} = \frac{\bar{c}^{(k)}_{\text{max}}}{n} ,
\ee
where we used $\sign(w^{(k)}_{i^*}) = \sign(c^{(k)}_{i^*})$, see \ref{sec:enter_the_active_set}.

\section{Implementation of the algorithms}
\label{sec:implementation}

\subsection{CD}
We implement CD in \texttt{Python} using \texttt{numpy version 2.2.2}.
We note that CD is also implemented in the subroutine \texttt{Lasso} in \texttt{scikit-learn version 1.6.1}.

\subsection{\LARS{} and \LARSLASSO{}}

\begin{algorithm}
\caption{Modified \textit{Least Angle Regression} (\LARSLASSO{})}\label{alg:LARSLASSO}
\begin{algorithmic}
\State Given $\bfX$ and $\bfy$
\State Set $k=0$, $\bfw^{(0)}=\bf0$, $\tilde i = -1$
\State $\bfy_{\parallel}=\bfX\left[\bfX^T\bfX\right]^{-1}\bfX^T\bfy$
\State $\bfc^{(0)} = \bfX^T \bfy_{\parallel}$
\State $\alpha^{(0)} = \bar{c}^{(0)}_{\text{max}} / n$
\While{$\|\bfw^{(k)}\|_0 < m-1$}
\State $\mathcal{A}^{(k)} = \left\{ i^*\in \{1,\dots,m\} \setminus \{\tilde i\} ~ \big| ~ |c^{(k)}_{i^*}| = \bar{c}^{(k)}_{\text{max}} = \max_i |c^{(k)}_i| \right\}$, $\mathcal{A}^{\complement(k)} = \left\{ j^* \in \{1,\dots,m\} ~ \big| ~ j^* \notin \mathcal{A}^{(k)} \right\}$
\State $\bfs^{(k)}=\sign(\bfc^{(k)})$
\State $\bar{\bfX}^{(k)}_{\mathcal{A}} = \bfX_{\mathcal{A}} \diag(\bfs^{(k)}_{\mathcal{A}})$
\State $A^{(k)} = 1/\sqrt{\boldsymbol{1}^T_{\mathcal{A}}\left[\bar{\bfX}^{T(k)}_{\mathcal{A}}\bar{\bfX}^{(k)}_{\mathcal{A}}\right]^{-1}\boldsymbol{1}_{\mathcal{A}}}$
\State $\bfu^{(k)} = A^{(k)} \bar{\bfX}^{(k)}_{\mathcal{A}}\left[\bar{\bfX}^{T(k)}_{\mathcal{A}}\bar{\bfX}^{(k)}_{\mathcal{A}}\right]^{-1}\boldsymbol{1}_{\mathcal{A}}$
\State $\gamma^{(k)} \gets \min_{j^* \in \mathcal{A}^{\complement(k)}}^+ \left\{
\frac{\bar{c}^{(k)}_{\text{max}} - c^{(k)}_{j^*}}{A^{(k)} - a^{(k)}_{j^*}},
\frac{\bar{c}^{(k)}_{\text{max}} + c^{(k)}_{j^*}}{A^{(k)} + a^{(k)}_{j^*}}
\right\}$
\State $\bfd^{(k)}_{\mathcal{A}} = A^{(k)} \diag(\bfs^{(k)}_{\mathcal{A}})\left[\bar{\bfX}^{T(k)}_{\mathcal{A}}\bar{\bfX}^{(k)}_{\mathcal{A}}\right]^{-1}\boldsymbol{1}_{\mathcal{A}}$
\State $\tilde\gamma \gets \min_{i^*\in\mathcal{A}^{(k)}}^+ \{ - w^{(k)}_{i^*}/d^{(k)}_{i^*} \}$
\If{$\{ - w^{(k)}_{i^*}/d^{(k)}_{i^*} \} \cap \Rset^+ \neq\emptyset$ and $\tilde\gamma < \gamma^{(k)}$}
    \State $\gamma^{(k)} \gets \tilde\gamma$
    \State $\tilde i \gets \argmin_{i^*\in\mathcal{A}^{(k)}}^+ \{ - w^{(k)}_{i^*}/d^{(k)}_{i^*} \}$ 
\Else
    \State $\tilde i \gets -1$ 
\EndIf
\State $\bfw^{(k+1)}_{\mathcal{A}^{\complement}} \gets \bfw^{(k)}_{\mathcal{A}^{\complement}}$
\State $\bfw^{(k+1)}_{\mathcal{A}} \gets \bfw^{(k)}_{\mathcal{A}} + \gamma^{(k)} A^{(k)} \diag(\bfs^{(k)}_{\mathcal{A}})\left[\bar{\bfX}^{T(k)}_{\mathcal{A}}\bar{\bfX}^{(k)}_{\mathcal{A}}\right]^{-1}\boldsymbol{1}_{\mathcal{A}}$
\State $\bfc^{(k+1)} = \bfX^T [\bfy_{\parallel} - \bfX\bfw^{(k+1)}]$
\State $\alpha^{(k+1)} = \bar{c}^{(k+1)}_{\text{max}} / n$
\State $k \gets k+1$
\EndWhile
\State $\bfw^{(k+1)} \gets \left[\bfX^T\bfX\right]^{-1}\bfX^T\bfy$
\State $\alpha^{(k+1)} = 0$
\end{algorithmic}
\end{algorithm}

\cref{alg:LARSLASSO} provides a detailed description of \LARSLASSO{}.
We implement \LARS{} and \LARSLASSO{} in \texttt{Python} using \texttt{numpy version 2.2.2}.
We note that \LARS{} and \LARSLASSO{} are also implemented in the subroutine \texttt{lars\_path} in \texttt{scikit-learn version 1.6.1}.


\subsubsection{Evaluating the equality of numbers}
An important question is how to evaluate numerically whether two numbers are equal, or equivalently how to evaluate whether a number is zero.
For example, in \cref{alg:LARS,alg:LARSLASSO}, the active set $\mathcal{A}^{(k)}$ is determined by identifying the maximum absolute correlation and checking which feature vectors correspond to an equal correlation in magnitude.
Further, when computing the $L_0$-pseudo-norm $\|\bfw^{(k)}\|_0$ in \cref{alg:LARS,alg:LARSLASSO}, we have to determine whether a number is zero.
To avoid the influence of small numerical variations, we define a tolerance value.
If the absolute difference between two numbers is below the tolerance, we claim that these numbers are equal.
Equivalently, if the absolute value of a number is below the tolerance, we claim that this number is zero.
In the code, we choose this tolerance as $10^{-12}$.

\subsubsection{Early stopping}
\label{sec:early_stopping}
\LARS{} and \LARSLASSO{}, see \cref{alg:LARS,alg:LARSLASSO}, can exhibit unpredictable behavior as the active set increases.
Particularly, solving the linear system $\left[\bar{\bfX}^{T(k)}_{\mathcal{A}}\bar{\bfX}^{(k)}_{\mathcal{A}}\right]^{-1}\boldsymbol{1}_{\mathcal{A}}$ may become ill-conditioned.
Following the implementation of the subroutine \texttt{lars\_path} in \texttt{scikit-learn version 1.6.1}, we stop the iterations once $\alpha=\bar{c}_{\text{max}} / n$ is below a threshold of \texttt{np.finfo(np.float32).eps}.

\subsection{ISTA}
We implement ISTA and pathwise ISTA in \texttt{Python} using \texttt{numpy version 2.2.2} and \texttt{PyTorch}, i.e., \texttt{torch version 2.6.0}.
Specifically, we use \texttt{PyTorch} to efficiently compute the gradient of $f(\bfw)$ using automatic differentiation.


\bibliographystyle{elsarticle-harv}
\bibliography{ALL}

\end{document}